\documentclass[12pt,preprint]{emulateapj}
\usepackage{natbib}
\usepackage{color}
\usepackage{hyperref} %----set up hyper link to citations, equations...
\hypersetup{colorlinks=true,citecolor=blue}

\def \bea {\begin{eqnarray}}
\def \ena {\end{eqnarray}}                  
\def    \simlt  {\lower.5ex\hbox{$\; \buildrel < \over \sim \;$}}
\def    \simgt  {\lower.5ex\hbox{$\; \buildrel > \over \sim \;$}}

\newcommand     \mum    {\,\mu{\rm m}}  % to use in math mode

\def	\cm		{\,{\rm {cm}}}

\def	\B		{{\rm B}}

\def	\erg		{\,{\rm {erg}}}

\def    \exp 		{\,{\rm {exp}}}
\def	\g		{\,{\rm g}}

\def	\Jy		{\,{\rm {Jy}}}

\def	\K		{\,{\rm K}}
\def    \kB    		{k_{\rm B}}

\def	\s		{\,{\rm s}}
\def	\sr		{\,{\rm {sr}}}
\def    \ln  		{\,{\rm {ln}}}
\def    \yr  		{\,{\rm {yr}}}

\def	\md		{\rm md}

\def	\H		{\rm H}

\def	\xhat		{\hat{\bf x}}
\def	\yhat		{\hat{\bf y}}
\def	\zhat		{\hat{\bf z}}
\def	\ahat		{\hat{\bf a}}
\def	\ehat		{\hat{\bf e}}

\def	\hhat		{\hat {\bf h}}

\def    \Bv     	{\bf  B}

\def    \rv     	{\bf  r}

\def	\ba			{{\bf a}}
\def	\be			{{\bf e}}

\def	\bB		{{\bf B}}

\def 	\bE		{{\bf E}}

\def	\bH		{{\bf H}}

\def	\bJ		{{\bf J}}

\def	\bM		{{\bf M}}

\def	\br		{{\rv}}
\def	\bv		{{\bf v}}

\def	\gas		{\rm {gas}}
\def	\th			{\rm {th}}

\def	\d			{\rm d}

\def    \abs     	{\rm {abs}}

\def    \ext    	{\rm {ext}}
\def    \pol    	{\rm {pol}}

\def	\eff		{\rm {eff}}

\def	\mag 	{\rm {m}}

\def	\tot		{\rm {tot}}

\def	\sp			{\rm {sp}}

\def    \Planck		{{\it Planck}~}

\def	\ed			{\rm {ed}}

\def	\cl		{\rm {cl}}

\font\mib=cmmib10

\def\bmu{\hbox{\mib\char"16}}

\def\bomega{\hbox{\mib\char"21}}

\begin{document}
\shortauthors{Hoang \& Lazarian}
\title{Polarization of Magnetic Dipole Emission and spinning dust emission from magnetic nanoparticles}

\author{Thiem Hoang\altaffilmark{1,2}, Alex Lazarian\altaffilmark{3}}
\altaffiltext{1}
{Institute of Theoretical Physics, Goethe  Universit$\ddot{\rm a}$t Frankfurt, D-60438 Frankfurt am Main, Germany}
\altaffiltext{2}
{Current address: Canadian Institute for Theoretical Astrophysics, University of Toronto, 60 St. George Street, Toronto, ON M5S 3H8, Canada}
\altaffiltext{3}
{
Department of Astronomy, University of Wisconsin-Madison
}

\begin{abstract}

Magnetic dipole emission (MDE) from interstellar magnetic nanoparticles is an important Galactic foreground in the microwave frequencies, and its polarization level may pose great challenges for achieving reliable measurements of cosmic microwave background (CMB) B-mode signal. To obtain theoretical constraints on the polarization of MDE, we first compute the degree of alignment of big silicate grains incorporated with magnetic inclusions. We find that, in realistic conditions of the interstellar medium, thermally rotating big grains with magnetic inclusions are weakly aligned and achieve {\it alignment saturation} when the magnetic alignment rate becomes much faster than the rotational damping rate. We then compute the degree of alignment for free-flying magnetic nanoparticles, taking into account various interaction processes of grains with the ambient gas and radiation field, including neutral collisions, ion collisions, and infrared emission. We find that the rotational damping by infrared emission can significantly decrease the degree of alignment of small particles from the saturation level, whereas the excitation by ion collisions can enhance the alignment of ultrasmall particles. Using the computed degrees of alignment, we predict the polarization level of MDE from free-flying magnetic nanoparticles to be rather low. Such a polarization level is within the upper limits measured for anomalous microwave emission (AME), which indicates that MDE from free-flying iron particles may not be ruled out as a source of AME.  We also quantify spinning dust emission from free-flying iron nanoparticles with permanent magnetic moments and find that its emissivity is one order of magnitude lower than that from spinning polycyclic aromatic hydrocarbons (PAHs). Finally, we compute the polarization spectra of spinning dust emission from PAHs for the different interstellar magnetic fields.

\end{abstract}


\section{Introduction}

Cosmic microwave background (CMB) experiments (see \citealt{Bouchet:1999p4616}; \citealt{Tegmark:2000p5597}; \citealt{Efstathiou:2003p4793}; \citealt{Bennett:2003p4582}) are of great importance for studying the early universe and its subsequent expansion. The era of precision cosmology with {\it Wilkinson Microwave  Anisotropy Probe} ({\it WMAP}) and \Planck requires an accurate model of Galactic foregrounds to allow the reliable subtraction of foreground contamination from the CMB radiation. 

Anomalous microwave emission (AME) in the $\sim$ 10--60 GHz frequency range is a new, important Galactic foreground component, which was discovered about 20 years ago (\citealt{Kogut:1996p5293}; \citealt{Leitch:1997p7359}). It appears that electric dipole emission (\citealt{1998ApJ...508..157D}, hereafter DL98) from rapidly spinning ultrasmall grains (mostly polycyclic aromatic hydrocarbons, hereafter PAHs; \citealt{1984A&A...137L...5L}; \citealt{1992A&A...253..498R}) is the most probable origin of AME (see \citealt{Collaboration:2013vx}; \citealt{PlanckCollaboration:2011p515}). Moreover, magnetic dipole emission (hereafter MDE) from magnetic particles was also suggested as a possible source of AME (\citealt{1999ApJ...512..740D}, hereafter DL99). \cite{2013ApJ...765..159D} (henceforth DH13) refined the MDE model by using the Gilbert equation approach to describe magnetization dynamics. With the revised magnetic susceptibility, DH13 show that MDE is an important foreground for frequencies $\nu=70-300$ GHz and contributes little to AME unless magnetic nanoparticles are extremely elongated. 

Ongoing and future CMB experiments aiming to detect primordial gravitational waves through B-mode polarization face great challenges from polarized Galactic foregrounds (\citealt{Ade:2015ee}; \citealt{2014arXiv1409.5738P}). Due to strong polarization of thermal dust emission above $\nu\sim 100$ GHz, the lower frequency range is expected to be a favored window for CMB B-mode measurements. In this frequency range, spinning dust emission appears to be weakly polarized (\citealt{2000ApJ...536L..15L}; \citealt{2013ApJ...779..152H} for theory, and \citealt{2015arXiv150606660P} for observations). {The remaining question is to what extent MDE from magnetic particles is polarized. Answering this question is not only essential for analysis of vast data from CMB B-mode experiments, but also provides unique insight into the origin of AME. } 

Previous works on the MDE polarization either assumed perfect alignment (DL99) or considered an arbitrary range of the degree of alignment (DH13) of magnetic particles with the magnetic field. { With the assumption of perfect alignment, MDE is predicted to have a high polarization level up to $30\%$, for which it is usually referred to distinguish MDE from spinning dust emission. While big grains with magnetic inclusions that rotate at suprathermal velocities due to radiative torques (\citealt{1996ApJ...470..551D}; \citealt{2009ApJ...695.1457H}) or/and Purcell torques \citep{1979ApJ...231..404P} can be perfectly aligned (\citealt{Lazarian:2008fw}), it is not the case for thermally rotating ones and free-flying magnetic nanoparticles.} To obtain reliable predictions for the MDE polarization, in this paper, we are going to numerically compute the degree of magnetic alignment for big grains with magnetic inclusions and for free-flying magnetic nanoparticles, using realistic magnetic susceptibility and accounting for various interaction processes (e.g., gas-grain collisions, ion-grain collisions, and infrared emission) (DL98; HDL10, hereafter HDL10).

In particular, magnetic nanoparticles have permanent magnetic moments due to spontaneous magnetization in the absence of external magnetic fields (see the next section). Thus, such magnetic particles while spinning certainly emit {\it rotational radiation}, which is essentially similar to rotational radiation from electric dipole  (\citealt{Erickson:1957p4806}). This effect will be quantified in our paper.

The paper is structured as follows. In Section \ref{sec:magrev} we briefly describe magnetic materials and summarize the principal formula of magnetic susceptibility to be used for later calculations of grain alignment. In Section \ref{sec:rotdamp} and \ref{sec:DG} we briefly discuss rotational damping and excitation for magnetic particles and magnetic alignment. Section \ref{sec:num} is devoted to present numerical methods and obtained results of magnetic alignment. In Section \ref{sec:mdepol}, we present the polarization of MDE predicted with computed degrees of alignment of magnetic particles. Spinning dust emission from magnetic particles is presented in Section \ref{sec:spinem}. An extended discussion and summary are presented in Section \ref{sec:discus} and \ref{sec:sum}, respectively.

\section{Magnetic Properties of Interstellar Dust}\label{sec:magrev}
\subsection{Classification of magnetic materials}
Iron is among the most abundant elements in the interstellar medium (ISM). Having four parallel electron spins in the $3d^{6}4s^{2}$ electronic shell, an Fe atom owns a large intrinsic magnetic moment. In solids, the effective magnetic moment per iron atom was well measured to be $\mu_{0}=p\mu_{B}$, where $\mu_{B}\equiv e\hbar/2m_e c$ is the Bohr magneton, and $p\sim 3$ for iron clusters of $\mathcal{N}<120$ atoms and $p\sim 2.2$ for $\mathcal{N}\sim 520$ \citep{1994Sci...265.1682B}. In the ISM, iron atoms may exist in the form of free-flying nanoparticles and/or are incorporated into big grains (either diffuse or cluster distribution), which produce various magnetic materials.

Magnetic materials are classified into the following types: paramagnetic, ferromagnetic (metallic iron), ferrimagnetic (e.g., Fe$_{3}$O$_{4}$, $\gamma$Fe$_{2}$O$_{3}$), and anti-ferromagnetic (e.g., Fe$_{2}$SiO$_{4}$) materials. Three latter materials, containing a high fraction of iron, exhibit magnetic ordering due to the exchange interaction between magnetic dipoles. Thus, throughout this paper, particles of these iron-rich materials are referred to as {\it magnetic} particles, and those of poor-iron materials (e.g., paramagnetic material) are referred to as {\it nonmagnetic} particles. Readers are referred to DL99 and DH13 for extended discussion on magnetic properties of interstellar dust. In the following, we briefly present essential contents needed for our discussion and calculations.

Based on magnetic configuration, iron-rich materials may exist in {single-domain and multiple-domain}. Magnetic particles with radius smaller than $a\sim 0.02\mum$ is usually single-domain magnetic material, while larger grains (e.g., bulk iron) develop multiple magnetic domains as a ways to achieve minimum magnetostatic energy for the system. \footnote{Throughout this paper, $a$ denotes the radius of spherical particles or the radius of an equivalent sphere of the same volume in the case of non-spherical particles.}%The spontaneous magnetization in the single-domain magnetic material is uniform.

{\bf Ferromagnetic and superparamagnetic materials:} Iron particles at low temperatures tend to be ferromagnetic in which the exchange interaction between electron spins (magnetic dipoles) enables them to be spontaneously aligned in an easy axis (e.g, long axis), resulting in spontaneous magnetization $M_{s}$. When the grain temperature exceeds the {\bf blocking} threshold $T_{b}$, the particle becomes superparamagnetic because thermal fluctuations of the lattice exceed the energy barrier (i.e., magnetostatic energy) to randomize the magnetizations, resulting in a zero net intrinsic magnetic moment. Experimental measurements (e.g., \citealt{Carvell:2010de}) show $T_{b}\sim 100$K for the $a\sim 3$ nm iron particles, whereas tiny iron grains of $\sim 20$ atoms are found to have nonzero magnetic moments at grain temperature $T_{d}\sim 120$ K \citep{1994Sci...265.1682B}. Therefore, for the ISM conditions, single-domain iron particles are essentially ferromagnetic due to low temperatures (i.e., $T_{d}<100$ K), probably, except during thermal spikes following absorption of ultraviolet photons. 

{\bf Ferrimagnetic and anti-ferromagnetic materials:} Structures Fe$_{3}$O$_{4}$ (magnetite) and $\gamma$ Fe$_{2}$O$_{3}$ (maghemite) exhibit ferrimagnetic behavior, having a nonzero spontaneous magnetization in the absence of external magnetic fields. Structures Fe$_{2}$SiO$_{4}$ exhibit anti-ferromagnetic behavior, in which the total spontaneous magnetization is zero because magnetizations are aligned in the exactly opposite directions due to the exchange interaction.

\subsection{Magnetic susceptibility}
For magnetic materials, the response of the magnetization ${\bf M}$ to an oscillating magnetic field is described by dynamical equations. To model the magnetic response of the ferromagnetic material, DL99 employed the Drude model with scalar susceptibility, in which the magnetic response to external field $\bH$ is described by
\bea
\ddot{{\bf M}}=\omega_{0}^{2}\left[\chi(0){\bf H}- {\bf M} \right]- \frac{\dot{{\bf M}}}{\tau_{0}},\label{eq:ddMdt}
\ena
where $\chi(0)$ is the magnetic susceptibility at zero frequency, $\omega_{0}$ is the resonance frequency, and $\tau_{0}$ is the magnetization damping time. 

Let ${\bf H}={\bf H}_{0} + {\bf h}e^{i\omega t}$, where $H_{0}$ is the static magnetic field and ${\bf h}e^{i\omega t}$ is the varying applied magnetic field. We can also write ${\bf M}= \bM_{s} + {\bf m}e^{i\omega t}$ where the second term denotes the varying magnetization. Then, $\dot{\bf M}=i\omega {\bf M}$, $\ddot{\bf M}=-\omega^{2} {\bf M}$. Plugging these expressions into Equation (\ref{eq:ddMdt}), one readily obtains
\bea
\chi(\omega)=\frac{\omega_{0}^{2}\chi(0)}{\omega^{2}-\omega_{0}^{2}-i\omega \tau_{0}^{-1}}=\frac{\chi(0)}{1-(\omega/\omega_{0})^{2}-i\omega \tau},\label{eq:drude}
\ena
where $\tau = (\omega_{0}^{2}\tau_{0})^{-1}$, which is the Drude complex susceptibility.

By multiplying the denominator of Equation (\ref{eq:drude}) with $1-(\omega/\omega_{0})^{2}+i\omega \tau$ and representing $\chi(\omega)=\chi_{1}(\omega) + i\chi_{2}(\omega)$, we obtain
\bea
\chi_{1}(\omega)=\frac{\chi(0)\left[1-(\omega/\omega_{0})^{2}\right]}{[1-(\omega/\omega_{0})^{2}]^{2}+(\omega \tau)^{2}},\\
\chi_{2}(\omega)=\frac{\chi(0)\omega\tau}{[1-(\omega/\omega_{0})^{2}]^{2}+(\omega \tau)^{2}}.\label{eq:chi2}
\ena

Assuming the critically-damped condition with $\omega_{0}\tau=2$, the above equations become
\bea
\chi_{1}^{cd}(\omega)&=&\frac{\chi(0) \omega \tau\left[1-(\omega\tau/2)^{2}\right]}{[1+(\omega \tau/2)^{2}]^{2}}\label{eq:chi1_cd},\\
\chi_{2}^{cd}(\omega)&=&\frac{\chi(0) \omega \tau}{[1+(\omega \tau/2)^{2}]^{2}}\label{eq:chi_cd},
%K(\omega)= \frac{\chi_{2}^{cd}(\omega)}{\omega}=\frac{\chi(0)\tau}{[1+(\omega \tau/2)^{2}]^{2}},\label{eq:Kappa_cd1}
\ena
where the denominator indicates that the magnetic susceptibility is significantly suppressed when the frequency of the applied field becomes larger than the response rate of the system (i.e., $\omega > \omega_{0}=2/\tau$).

% (metallic Fe, ferri, anti-ferri = magnetic particles. Metallic Fe particle consists of only Fe in bcc=body-centered cubic structure
\subsection{Free-flying single-domain magnetic nanoparticles}

For single-domain magnetic particles with nonzero spontaneous magnetization $M_{s}$, only the magnetic field component perpendicular to ${\bf M}_{s}$ can cause the reorientation of ${\bf M}_s$, producing anisotropic magnetic susceptibility, which is characterized by $\chi_{\|}(0)=0$ and $\chi_{\perp}(0)$. For three representative structures of magnetic particles, metallic iron (Fe), maghemite (Fe$_{3}$O$_{4}$) and magnetite ($\gamma$Fe$_{2}$O$_{3}$), which have $4\pi M_{s}=22000, 6400,$ and $4780$G, respectively (see Table 1 in DL99), the numeric estimates in DL99 give $\chi_{\perp}(0)=3.3, 0.83, 0.6$ for Fe, Fe$_{3}$O$_{4}$, and $\gamma$Fe$_{2}$O$_{3}$, respectively.

The complex magnetic susceptibility $\chi_{\perp}(\omega)$ is given by Equations (\ref{eq:chi1_cd}) and (\ref{eq:chi_cd}) where $\chi(0)$ is replaced by $\chi_{\perp}(0)$,
%\bea
%\chi_{\perp,2}(\omega)=\frac{\chi_{\perp}(0)\omega \tau}{[1+(\omega \tau/2)^{2}]^{2}}.\label{eq:chiiron}
%\ena
and the value of $\tau$ is evaluated based on the Larmor precession frequency $\omega_{g}$ of the atomic magnetic moment $\mu_{0}$ around the internal magnetic field. For single-domain magnetic particles, the internal field is uniform of $4\pi M_{s}/3$, which yields $\omega_{g} = (2e/2m_{e}c)4\pi M_{s}/3$ with $M_{s}$ measured in units of G. For metallic Fe, one obtains $\omega_{g}/2\pi = 20.51$ GHz and $\tau=2/\omega_{g} \approx 1.6\times 10^{-11}\s$. Similarly, $\tau\approx 5.3\times 10^{-11}\s$ and $7.1\times 10^{-11}$s for Fe$_{3}$O$_{4}$, and $\gamma$Fe$_{2}$O$_{3}$.

DH13 employed the Gilbert equation \citep{Gilbert:2004gx} in which the magnetization damping is characterized by a dimensionless Gilbert parameter $\alpha_{G}$, and derived the following susceptibility for single-domain ferromagnetic material:
\bea
\chi_{1,\pm}(\omega)&=&\frac{\omega_{M}(\omega_{0}\mp \omega)}{(\omega_{0}\mp \omega)^{2}+\alpha_{G}^{2}\omega^{2}},\label{eq:chi1_DH}\\
\chi_{2,\pm}(\omega)&=&\frac{\alpha_{G}\omega_{M}\omega}{(\omega_{0}\mp \omega)^{2}+\alpha_{G}^{2}\omega^{2}},\label{eq:chi2_DH}
\ena
where the plus and minus sign corresponds to the applied field ${\bf h}e^{i\omega t}$ rotating anticlockwise and clockwise around ${\bf M}_{s}$, $\omega_{0}= 32$GHz and $\omega_{M}/2\pi=4.91$GHz for Fe prolate spheroid of axis ratio $2:1$ (see Table 2 in DH13 for more detail).

{The susceptibility of ferromagnetic single-domain (SD) particles from DL99 and the average susceptibility ($\chi_{2}=\chi_{2,+}+\chi_{2,-})/2$) from DH13 with $\alpha_{G}=0.2$ are shown in Figure \ref{fig:chi_omega} (see orange lines). For frequencies $\nu<5$ GHz relevant for grain alignment under interest in this paper, two models are only different by a numeric factor, and become identical when $\alpha_{G}$ is increased by a factor of $112$.}

\subsection{Big grains with magnetic inclusions}
The incorporation of magnetic particles with $N_{\cl}$ iron atoms per particle into a big nonmagnetic grain is expected to greatly enhance the grain magnetic susceptibility.\footnote{For Fe with mass density $\rho=7.86\g\cm^{-3}$, $N_{\cl}=4\pi/3 a^{3}\rho/m_{Fe}\simeq 355a_{-7}^{3}$ with $a_{-7}=a/10^{-7}\cm$.}  Although the possibility of having magnetic inclusions in big carbonaceous grains may not be ruled out,\footnote{Yet, if it happens, big carbon grains with inclusions would be aligned with magnetic fields, which is currently not observed.} in this paper, we only consider the case of magnetic inclusions in big silicate grains.

%Inclusion of iron nanoparticles. Note that ferromagnetic particle = iron nanoparticles are single-domain ferromagnetic with Ms
\subsubsection{Single-domain ferromagnetic inclusions}
Let us assume that single-domain ferromagnetic inclusions are randomly distributed in a big grain, and that each inclusion is spontaneously magnetized along its easy-axis. In an applied field $\bH$, $\sim $ 2/3 of inclusions have their spontaneous magnetizations $\bM_{s}$ perpendicular to $\bH$, which contribute to the susceptibility. And $\sim $ 1/3 of inclusions have $\bM_{s}$ parallel to $\bH$, which do not contribute to the susceptibility (DL99). The zero-frequency effective susceptibility per unit volume of the composite grain is given by
\bea
\chi_{\eff}(0) = \frac{2\phi\chi_{\perp}(0)/3}{1+(4\pi/3)\chi_{\perp}(0)[1-2\phi/3]},\label{eq:chi_ferroinc}
\ena
and $\chi_{\eff}(\omega)$ is then evaluated by Equation (\ref{eq:chi_cd}) where $\chi(0)$ is substituted with $\chi_{\eff}(0)$ (see DL99).

\subsubsection{Superparamagnetic inclusions}

When ferromagnetic inclusions are sufficiently small so that thermal energy can exceed the energy barrier to reorient the inclusion magnetic moments, then thermal fluctuations within the grain can induce considerable fluctuations of the inclusion magnetic moments. In thermal equilibrium, the average magnetic moment of the ensemble of magnetic inclusions can be described by the Langevin function with argument $m H/k_{\B}T_{d}$, where $m= N_{\cl}\mu_{0}\equiv M_{s}V$ is the total magnetic moment of the cluster, and $H$ is the applied magnetic field (\citealt{Jones:1967p2924}; henceforth JS67). The composite grain exhibits superparamagnetic behavior, which has the magnetic susceptibility given by
\bea
\chi_{\sp}(0)=\frac{n_{\cl}m^{2}}{3k_{\B}T_{d}},\label{eq:chisp_rest}
\ena
where $n_{\cl}$ is the number of iron clusters per unit volume (JS67). 

Plugging $m$ and $n_{cl}=\mathcal{N}_{cl}/V$ with the total number of clusters $\mathcal{N}_{cl}=3.5\times 10^{8}\phi_{sp}N_{cl}^{-1}a_{-5}^{3}$ into Equation (\ref{eq:chisp_rest}), one obtains:
\bea
\chi_{\sp}(0)\approx 0.035N_{cl}\phi_{\sp}(p/5.5)^{2}\hat{T}_{d}^{-1},\label{eq:chi_superinc}
\ena
where $\phi_{\rm sp}$ is the volume filling factor, and $\hat{T}_{d}\equiv T_{d}/15\K$.

%If the total number of iron atoms incorporated in a grain is kept constant and the only change is from the diffuse distribution to clusters, then we have $N_{\rm Fe} \equiv n_{p}V = n_{\cl}VN_{\cl}$ where $n_{p}$ is the density of Fe atoms for the standard paramagnetic material. Therefore,
%\bea
%\chi_{\sp}(0)& =& n_{\cl}\frac{\left(N_{\cl}\mu_{0}\right)^{2}}{3k_{\B}T_{d}}=N_{\cl}\frac{n_{p}\mu_{0}^{2}}{3k_{\B}T_{d}},\nonumber\\
%&\approx & 1.24 \times 10^{-2} N_{\cl} f_p \hat{n}_{23} \hat{p}^2 \hat{T}_{d}^{-1},\label{eq:chi_superinc}
%\ena
%where $f_{p}=n_{p}/n$ with $n$ the atomic density is the fraction of paramagnetic (Fe) atoms, $\hat{p}=p/3$, $\hat{n}_{23}\equiv n/10^{23}\cm^{-3}$ and $\hat{T}_{d}\equiv T_{d}/15\K$. 

Superparamagnetic inclusions undergo thermally activated remagnetization at rate
\bea
\tau^{-1}_{\sp}\approx \nu_0 \exp\left[-N_{\cl}T_{\rm act}/T_{d}\right]
\label{sp}
\ena
where $\nu_0\approx 10^{9}\s^{-1}$ and $T_{\rm act}\approx 0.011\K$ (see \citealt{Morrish:2001vp}). The frequency dependence susceptibility $\chi_{\rm sp}(\omega)$ is then evaluated by Equation (\ref{eq:chi_cd}) where $\chi(0)=\chi_{\sp}(0)$ and $\tau=\tau_{\sp}$.

The total magnetic susceptibility of superparamagnetic grains now becomes (DL99)
\bea
\chi(\omega) = \chi_{\eff}(\omega) + \chi_{\sp}(\omega),\label{eq:chi_superinc}
\ena
where $\chi_{\sp}(\omega)$ is dominant for low frequencies of $\nu<5$ GHz, and $\chi_{\eff}(\omega)$ dominates for $\nu>5$ GHz (see Figure \ref{fig:chi_omega}, left panel).

\subsubsection{Effect of ferromagnetic inclusions on paramagnetic atoms}
Ferromagnetic inclusions in a big silicate grain generate static magnetic fields acting on nearby paramagnetic (iron) atoms, which is found to significantly increase the magnetic susceptibility of the composite system through spin-lattice relaxation (\citealt{1978ApJ...219L.129D}). The total magnetic susceptibility is equal to
\bea
\chi_{2}(\omega) = F\frac{\chi(0)\omega\tau_{1}}{1+(\omega \tau_{1})^{2}} + (1-F)\frac{\chi(0)\omega\tau_{2}}{1+(\omega \tau_{2})^{2}},\label{eq:K_ferropara}
\ena
where $\tau_{1}$ and $\tau_{2}$ are the spin-lattice and spin-spin relaxation times, and the susceptibility of the ordinary paramagnetic material $\chi(0)$ is
\bea
\chi(0)\simeq 0.042f_{p}\hat{n}_{23}(p/5.5)^{2}\hat{T}_{d}^{-1}
\ena
where $f_{p}=n_{p}/n$ with $n$ the atomic density is the fraction of paramagnetic (Fe) atoms, and $\hat{n}_{23}\equiv n/10^{23}\cm^{-3}$. 

The factor $F$ reads 
\bea
F=\frac{H_{0}^{2}}{H_{0}^{2}+H_{c}^{2}} {~\rm with~} H_{c}=\left(\frac{C_{M}T_{d}}{\chi(0)} \right)^{1/2},
\ena
where $C_{M}$ is the heat volume capacity at constant magnetization, $H_{c}\sim 10^{3}$ Oe for various paramagnetic materials, and $H_{0}$ is the rms value of the magnetic field in a given direction outside the ferromagnetic inclusions (see \citealt{Draine:1996p6977}).

Let $f_{F}$ be the fraction of the total atoms present in single-domain ferromagnetic inclusions. Then, 
\bea
H_{0}=3.8f_{F}np\mu_{B}/\sqrt{3}\simeq 112(f_{F}/0.01)(p/5.5) {\rm Oe}.
\ena

\begin{figure*}
\includegraphics[width=0.45\textwidth]{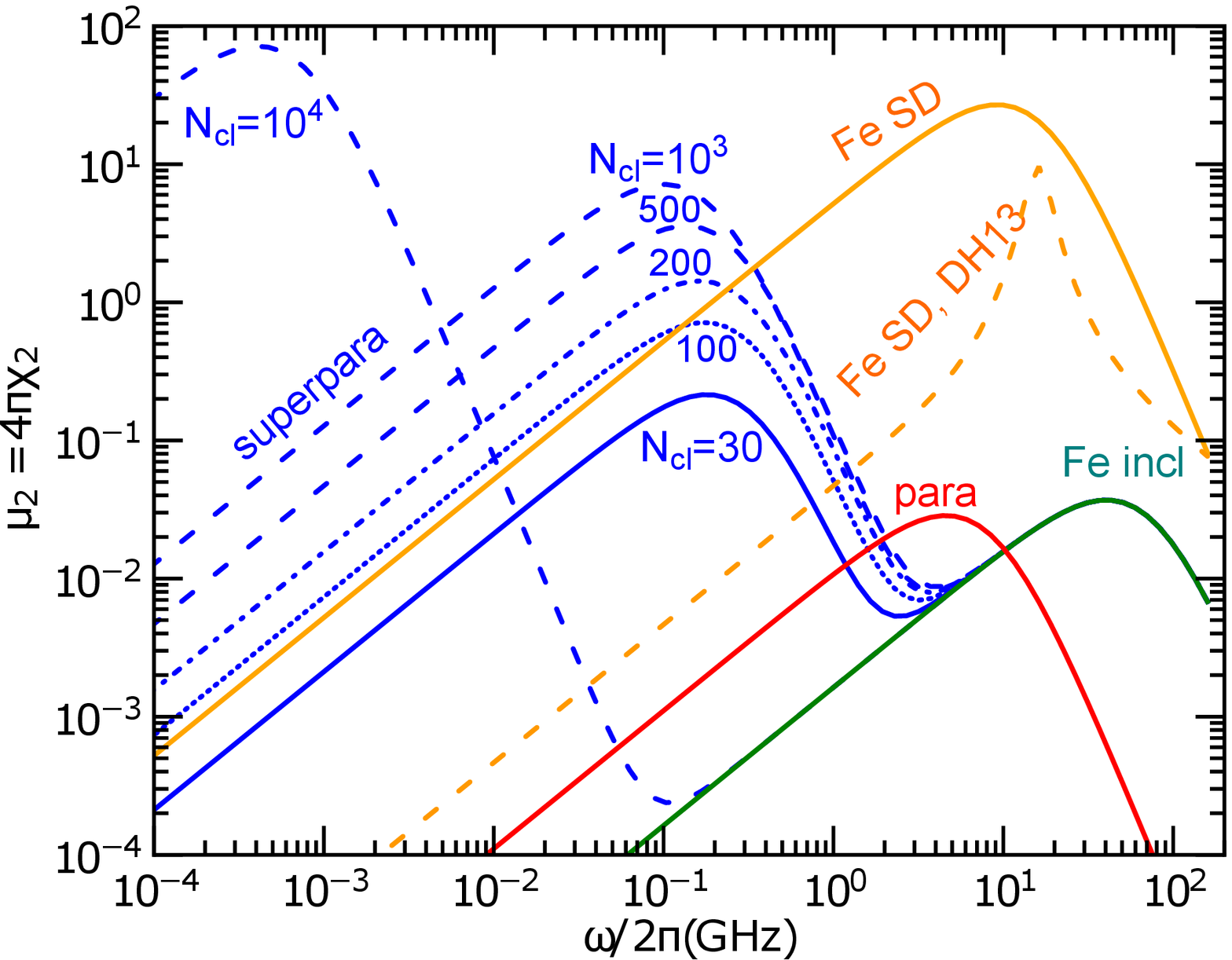}
\includegraphics[width=0.45\textwidth]{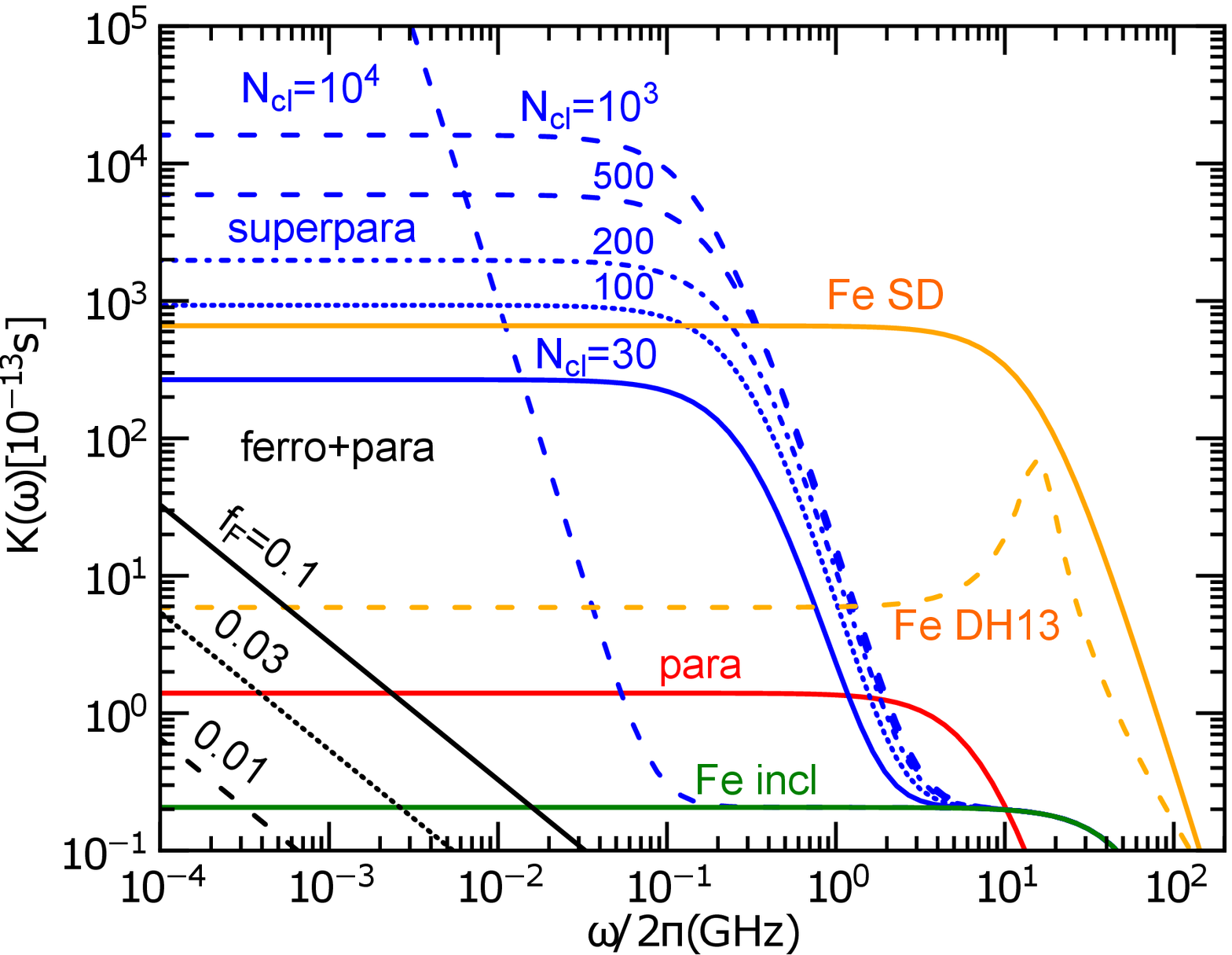}
\caption{Magnetic susceptibility (left panel) and $K(\omega)=\chi_{2}(\omega)/\omega$ (right panel) for different magnetic materials as functions of frequency. Typical grain temperature $T_{d}=18$K is assumed. The ferromagnetic susceptibility of DH13 is also shown for comparison. Superparamagnetic susceptibility is computed for $\phi_{\rm sp}=0.03$.}
\label{fig:chi_omega}
\end{figure*}

With the typical value of $\tau_{1}\sim 10^{-6}s$, the ferromagnetic-paramagnetic interaction can raise $\chi_{2}(\omega)$ by two orders of magnitude above the susceptibility of the ordinary paramagnetic material (see Figure \ref{fig:chi_omega}). 

Figure \ref{fig:chi_omega} also shows the frequency dependence of $\mu_{2}=4\pi \chi_{2}$ (left panel), and $K(\omega)=\chi_{2}(\omega)/\omega$ (right panel) for a variety of magnetic materials considered in this paper, including single-domain Fe nanoparticles (Fe SD), ferromagnetic and superparamagnetic inclusions with different $N_{\cl}$, and ferromagnetic-paramagnetic interactions. Results for the ordinary paramagnetic material ($f_{p}=0.1,p=5.5$) are also shown for comparison. Ferromagnetic susceptibility is the most important for high frequencies of $\nu>10$ GHz, which result in magnetic dipole emission (DL99; DH13). 

%Magnetic particles = ferro (iron), ferri, anti-ferri particles, not only iron particle. Higher impact than just iron particle
\section{Rotational damping and excitation for magnetic particles}\label{sec:rotdamp}
\subsection{Electric dipole and magnetic dipole moments}
Unlike PAHs, ferromagnetic particles do not possess intrinsic electric dipole moment because of the lack of polar bonds. In principle, any particles may have a net electric dipole moment when the charge distribution is asymmetric such that the charge centroid is displaced from the grain center of mass. However, due to high conductivity, magnetic particles cannot maintain such dipole moment because free electrons and holes can easily move in the grain (electron diffusion time $\tau_{diff}\sim 10^{-13}s$ much shorter than charging time) (see \citealt{deHeer:2009ta} for a review). 

At low temperatures, a magnetic particle of volume $V$ has an intrinsic magnetic moment due to spontaneous magnetization given by:
\bea
\mu_{\md}=M_{s}V=\frac{4\pi M_{s}}{3000\rm G} a_{-7}^{3}10^{-18} {\rm esu}
= \frac{4\pi M_{s}}{3000\rm G} a_{-7}^{3} {\rm D},~~~\label{eq:Ms}
\ena
where $V=4\pi a^{3}/3$ and 1 D $= 10^{-18}$ statC.cm. 

{Thus, iron particles of $4\pi M_{s}=22000$G have largest value of $\mu_{\md}\simeq 7.33a_{-7}^{3}$ D, slightly lower than the electric dipole moment of PAHs with $\mu_{\ed}\sim 9.3a_{-7}^{3/2}$ D at $a=1$ nm (see DL98). Note that rotational emission power by the magnetic dipole varies with grain size as $\mu_{\rm md}^{2}\omega^{4}\propto a_{-7}^{3}\omega^{4}$, instead of $a_{-7}^{3}\omega^{4}$ for the case of the electric dipole, which reveal that larger magnetic grains seem to emit more spinning radiation than the electric grains.}

\subsection{Damping and Excitation Coefficients}

Rotational damping and excitation for dust grains in general arise from collisions between grains and gaseous atoms followed by the evaporation of atoms/molecules from the grain surface, absorption of starlight and IR emission (DL98; HDL10). If the grain posses electric dipole moment, the distant interaction of the grain electric dipole with passing ions results in an additional effect, namely plasma drag. The damping and excitation for these processes are described by the dimensionless damping coefficient $F$ and excitation coefficient $G$, respectively (see Appendix \ref{apdx:FGcoeff}). 

It is well-known that, for UV-IR photons with $\nu>10^{3}$ GHz, the magnetic properties of dust grains can be ignored in calculations of absorption cross-section $C_{\abs}$ due to the dominance of electric dipole cross-section (see DL99; DH13). In this case, grain charging for magnetic grains may be treated by the similar model as silicate grains (see \citealt{2001ApJS..134..263W}), and collisional damping and excitations of magnetic grains are computed as in DL98. Damping and excitation by IR emission is also computed using the general approach in DL98, and the factor $2$ increase in the excitation coefficient is included (see HDL10).

For a magnetic grain, distant interactions with passing ions is also possible through the interaction of spontaneous magnetization with the magnetic field generated by moving ions in the grain's frame. However, this effect is negligibly small as shown in Appendix \ref{apd:plasma}.

Moreover, rotational emission by magnetic dipole moment results in the rotational damping. The characteristic damping time due to dipole emission $\tau_{\ed}$ is calculated with Equation (\ref{eq:tauedxy}) by replacing $\mu_{\ed}^{2}$ with $\mu_{\md}^{2}$, where the magnetic dipole is assumed to be in the plane perpendicular to the grain symmetry axis. We found that the rotational dipole damping is rather inefficient for magnetic nanoparticles.

\begin{figure}
\includegraphics[width=0.45\textwidth]{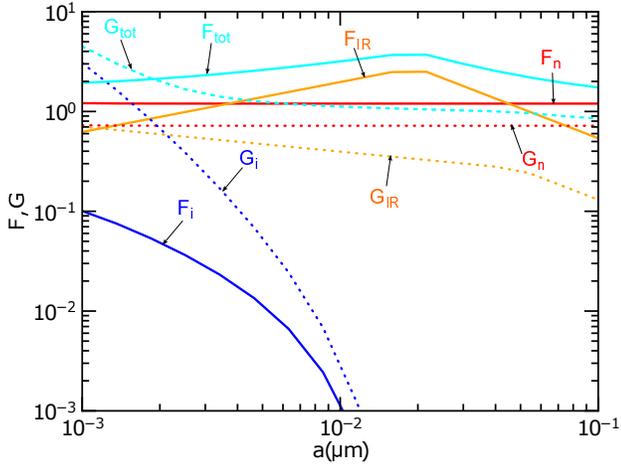}
\caption{Rotational damping ($F_{j}$; solid lines) and excitation ($G_{j}$; dotted lines) coefficients where $j=i, n, IR$ denotes the collision with incident ions, incident neutrals, and IR emission for ferromagnetic nanoparticles in the CNM.}
\label{fig:FG}
\end{figure}

Figure \ref{fig:FG} shows the values of $F$ and $G$ for various processes for ferromagnetic grains in the cold neutral medium (CNM; gas density $n_{\H}=30\cm^{-3}$ and temperature $T_{\gas}=100\K$) computed for an oblate spheroidal grain rotating along its symmetry axis. As shown, IR emission is dominant for rotational damping of small grains ($a<0.05\mum$), and neutral collisions are dominant for rotational damping/excitation for larger grains. Ion collisions dominate rotational excitation of ultrasmall grains in which the grains are negatively charged due to electron capture. 

\section{Magnetic Alignment of thermally rotating grains}\label{sec:DG}
\subsection{Davis-Greenstein Magnetic Relaxation}
%\subsubsection{Isotropic susceptibility}
\cite{1951ApJ...114..206D} (henceforth DG51) suggested that a paramagnetic grain rotating with angular velocity $\bomega$ in an external magnetic field experiences paramagnetic relaxation, which dissipates the grain rotational energy into heat. This results in the gradual alignment of $\bomega$ and $\bJ$ with the magnetic field at which the rotational energy is minimum.

Let us consider a spinning oblate spheroid with isotropic magnetic susceptibility. This isotropic susceptibility occurs in some materials, including big grains with randomly oriented magnetic inclusions, superparamagnetic grains. The rotating magnetic field component in the grain frame is decomposed into two components along $\ahat_{2}$ and $\ahat_{3}$, such that ${\bf B}_{\rm rot}=B_{\ext}\sin\beta (\cos\omega t \ahat_{2} + \sin\omega t \ahat_{3})$. The rotational energy dissipation is approximately given by
\bea
\frac{dW}{dt} = -\omega\chi_{2}(\omega)\langle VB_{rot}^{2}\rangle= -K(\omega)VB^{2}\omega^{2}\sin^{2}\beta,\label{eq:dWdt_iso}
\ena
where $K(\omega) = \chi_{2}(\omega)/ \omega$.

%\subsubsection{Anisotropic susceptibility}
Ferromagnetic and ferrimagnetic particles exhibit {\it anisotropic susceptibility} determined by the spontaneous magnetization ${\bf M}_{s}$. Assuming that the easy axis (i.e., ${\bf M}_{s}$) is along a long axis $\ahat_{3}$\footnote{This situation is idealized because oblate iron particles exhibit equatorial plane of saturation.}, and the grain is spinning along the short symmetry axis $\ahat_{1}$, then, the rotating magnetic field in the grain frame that results in energy dissipation is ${\bf B}_{\rm rot}=B_{\rm ext}\sin\beta \cos\omega t \ahat_{2}$.\footnote{For simplicity, previous studies of ferromagnetic alignment usually assumed prolate spheroids because the spontaneous magnetization is directed along the grain symmetry axis (\citealt{1958ApJ...128..497H}; \citealt{Jones:1967p2924}). Meanwhile, the Davis-Greenstein alignment is usually treated for oblate spheroids (\citealt{1997MNRAS.288..609L}; \citealt{1999MNRAS.305..615R}, henceforth RL99). Here, we consider oblate spheroid for the sake of consistency.} As a result, we have
\bea
\frac{dW}{dt} &=&- \frac{1}{2}\omega \chi_{\perp 2}(\omega)VB^{2}\sin^{2}\beta,\nonumber\\
& =&- K(\omega)VB^{2}\omega^{2}\sin^{2}\beta,\label{eq:dWdt_ani}
\ena
where $K(\omega) = \chi_{\perp, 2}(\omega)/(2\omega)$. 

%{\bf DH13 derived a more physical form of ferromagnetic susceptibility than DL99, but note that it is important only for high frequencies where MDE is crucial. For low frequencies where alignment is concerned, it leaves us with the DL99 susceptibility (see more detail in Section \ref{sec:DL_DH}).}

The characteristic timescale for the magnetic alignment of $\bJ$ with $\bB$ is obtained by setting $dW/dt=-d/dt(1/2I_{\|}\omega^{2}\sin^{2}\beta)$:
\bea
\tau_{\rm m} =\frac{I_{\|}}{K(\omega)VB^{2}}=\frac{2\rho a^{2}s^{-2/3}}{5K(\omega)B^{2}},\label{eq:tau_DG}
\ena
where $I_{\|}$ from Equation (\ref{eq:Iparperp}) has been used. 

To describe the effect of grain alignment by magnetic relaxation against the randomization by gas atoms, a dimensionless parameter $\delta_{m}$ is usually used:
\bea
\delta_{m}=\frac{\tau_{\H,\|}}{\tau_{\rm m}}\simeq 0.28\frac{ a_{-5}^{-1}s^{4/3}\hat{B}^{2}(K(\omega)/10^{-13}\s^{-1})}{\hat{n}_{\gas}\hat{T}_{\gas}\Gamma_{\|}},\label{dis3}
\ena
where $\tau_{\H,\|}$ is the gaseous damping for the rotation around the grain symmetry axis, $\hat{n}_{\gas}=n_{\H}/30\cm^{-3}$, $\hat{T}_{\gas}=T_{\gas}/100\K$, $\hat{B}=B/10\mu{\rm G}$, and $\Gamma_{\|}<1$ is a geometrical factor (see Appendix \ref{apdx:FGcoeff}). 

The $a\sim 0.1\mum$ grains have $\delta_{m}<1$ for the ordinary paramagnetic material, and $\delta_{m}\gg 1$ for the superparamagnetic and ferromagnetic materials (see Figure \ref{fig:chi_omega}). In the absence of suprathermal rotation, it follows that paramagnetic grains are likely randomized by gas collisions well before the magnetic relaxation, whereas magnetic relaxation rapidly brings superparamagnetic/ferromagnetic grains to be aligned with the magnetic field. However, the magnetic alignment for the latter materials is still not perfect, as shown in our next section.

\subsection{Resonance Magnetic Relaxation}
{Resonance magnetic relaxation has been shown to be important for the weak alignment of PAHs  (\citealt{2000ApJ...536L..15L}; \citealt{{2013ApJ...779..152H},{Hoang:2014cw}}, hereafter HLM13, HLM14), and it becomes dominant over the normal paramagnetic relaxation when the grain rotation frequency exceeds the spin-spin relaxation rate $\tau_{2}^{-1}$. For ferromagnetic particles, it is easy to see that resonance relaxation is unimportant because these particles have $\tau_{2}^{-1}\sim 5\times 10^{10}\s^{-1}$, which is still larger than the thermal rotation speed $\omega_{T} \simeq 3.3\times 10^{10}a_{-7}^{-5/2}s^{2/3}\hat{T}_{\gas}^{1/2}\s^{-1}$ (see Eq. \ref{eq:omegaT}) for tiny particles of $a\sim 1$nm.}

\subsection{Degrees of Grain Alignment}
Let $Q_{X}=\langle G_{X}\rangle $ with $G_{X}=\left(3\cos^{2}\theta-1\right)/2$ be the degree of internal alignment of the grain symmetry axis $\ahat_{1}$ with $\bJ$, and let $Q_{J}=\langle G_{J}\rangle$ with $G_{J}=\left(3\cos^{2}\beta-1\right)/2$ be the degree of external alignment of $\bJ$ with $\Bv$. Here $\theta$ is the angle between $\hat{\ba}_{1}$ and $\bJ$, and the angle brackets denote the average over the ensemble of grains. The net degree of alignment of $\ahat_{1}$ with $\Bv$, namely the Rayleigh reduction factor, is defined as $R = \langle G_{X}G_{J}\rangle$.

\section{Numerical Method and Results}\label{sec:num}
\subsection{Numerical Method}

As in previous works (RL99; HLM14), to study the alignment of the grain angular momentum $\bJ$ with the ambient magnetic field $\Bv$, we solve the Langevin equations for the evolution of $\bJ$ in time in an inertial coordinate system denoted by unit vectors $\ehat_{1}\ehat_{2}\ehat_{3}$ where $\ehat_{1}$ is chosen to be parallel to $\Bv$. The Langevin equations (LEs) read 
\bea
dJ_{i}=A_{i}dt+\sqrt{B_{ii}}dW_{i}\mbox{~for~} i=~1,~2,~3,\label{eq:dJ_dt}
\ena
where $dW_{i}$ are the random variables drawn from a normal distribution with zero mean and variance $\langle dW_{i}^{2}\rangle=dt$, and $A_{i}=\langle {\Delta J_{i}}/{\Delta t}\rangle$ and $B_{ii}=\langle \left({\Delta J_{i}}\right)^{2}/{\Delta t}\rangle$ are the drifting (damping) and diffusion coefficients defined in the $\ehat_{1}\ehat_{2}\ehat_{3}$ system. 

The drifting and diffusion coefficients in a frame of reference fixed to the grain body, $A_{i}^{b}$ and $B_{ij}^{b}$, are related to the damping and excitation coefficients ($F$ and $G$) as follows:
\bea
A_{i}^{b}&=&-\frac{J_{i}^{b}}{\tau_{\gas,i}}=-\frac{J_{i}^{b}}{\tau_{\H,i}}F_{\tot,i},\\
B_{11}^{b}&=&B_{\|}=\frac{2I_{\|}\kB T_{\rm gas}}{\tau_{\rm H,\|}}G_{\rm tot,\|},\\
B_{22}^{b}&=&B_{33}^{b}=B_{\perp}=\frac{2I_{\perp}\kB T_{\rm gas}}{\tau_{\rm H,\perp}}G_{\rm tot,\perp},
\ena
where  $F_{\tot,i}$ and $G_{\tot, ii}$ for $i=1,2,3$ (or $\|, \perp$) are the total damping and excitation coefficients from various processes, and $\tau_{\gas, i}=\tau_{\H,i}/F_{\tot,i}$.  Finally, $A_{i}$ and $B_{ii}$ are obtained by using the transformation of coordinate systems for $A_{i}^{b}, B_{ii}^{b}$ from $\ahat_{1}\ahat_{2}\ahat_{3}$ to  $\ehat_{1}\ehat_{2}\ehat_{3}$ (see HLD10; HLM14).

To incorporate magnetic relaxation, we need to add a damping term $-{J_{2,3}}/{\tau_{\mag}}$ to $A_{2,3}$ and an excitation term $B_{{\mag},22}=B_{{\mag},33}$ to $B_{22}$ and $B_{33}$, respectively. In the dimensionless units of $J'\equiv J/I_{\|}\omega_{T}$ and $t'\equiv t/\tau_{\H,\|}$, Equation (\ref{eq:dJ_dt}) becomes 
\bea
dJ'_{i}=A'_{i}dt'+\sqrt{B'_{ii}}dw'_{i} \mbox{~for~} i= 1,~2,~3,\label{eq:dJp_dt}
\ena
where $\langle dw_{i}^{'2}\rangle=dt'$ and
\bea
A'_{i}&=&-{J'_{i}}\left[\frac{1}{\tau'_{\gas,{\eff}}} +\delta_{m}(1-\delta_{1i})\right] -\frac{2}{3}\frac{J_{i}^{'3}}
{\tau'_{\ed,{\eff}}},\label{eq:Ai}~~~~\\
B'_{ii}&=&\frac{B_{ii}}{2I_{\|}\kB T_{\gas}}\tau_{\H,\|}+\frac{T_{\d}}{T_{\gas}}\delta_{\mag}(1-\delta_{1i}).\label{eq:Bii}
\ena
Above, $\delta_{1i}=1$ for $i=1$ and $\delta_{1i}=0$ for $i\ne 1$, and
\bea
\tau'_{\gas,{\eff}}= \frac{\tau_{\gas,{\eff}}}{\tau_{\H,\|}},~\tau'_{\ed,{\eff}}&=&\frac{\tau_{\ed,{\eff}}}{\tau_{\H,\|}},~~~
\ena
where $\tau_{\gas,{\eff}}$ and $\tau_{\ed,{\eff}}$ are the effective damping times due to dust-gas interactions and electric dipole emission (see Eq. E4 in HDL10). 

To numerically solve the Langevin equation (\ref{eq:dJp_dt}), earlier works (RL99; HLM14) usually used the first-order integrator, namely Euler-Maruyama algorithm. As shown in \cite{VandenEijnden:2006gp}, the second-order integrator is about one order of magnitude more accurate than the first-order one. For the second order, the angular momentum component $j_{i}\equiv J'_{i}$ at iterative step $n+1$ is evaluated as follows:
\bea
{j}_{i;n+1} &=& j_{i;n}  -  \gamma_{i}{j}_{i;n}h+\sqrt{h}\sigma_{ii}{\zeta}_{n}- \gamma_{i}\mathcal{A}_{i;n}-\gamma_{\ed}\mathcal{B}_{i;n},~~~
\ena
where $h$ is the timestep, $\gamma_{i}=1/\tau'_{\gas,{\eff}} +\delta_{m}(1-\delta_{zi})$, $\gamma_{\ed}=2/(3\tau'_{\ed,\eff})$, $\sigma_{ii}=\sqrt{B'_{ii}}$, and
\bea
\mathcal{A}_{i;n}&=& -\frac{h^{2}}{2} \gamma_{i} j_{i;n}+\sigma_{ii} h^{3/2}g(\xi_{n},\eta_{n})-\gamma_{\ed}j_{i;n}^{3}\frac{h^{2}}{2},\\
\mathcal{B}_{i;n}&=&j_{i;n}^{3}h - 3\gamma_{i}j_{i;n}^{3}\frac{h^{2}}{2}-\frac{3j_{i;n}^{5}\gamma_{\ed}h^{2}}{2}+3j_{i;n}^{2}\sigma_{ii} h^{3/2}g(\xi_{n},\eta_{n}),
\ena
with ${\eta}_{n}$ and ${\zeta}_{n}$ being independent Gaussian variables with zero mean and unit variance and $g(\xi_{n},\eta_{n})= \xi_{n}^{2} /2 + \eta_{n}^{2} /2\sqrt{3}$ (see Appendix \ref{apdx:LEsolver2} for details).

The timestep $h$ is chosen by $h= 0.01\min[1/F_{\tot,\|}, 1/G_{\tot,\|}, \tau_{\rm ed,\|}/\tau_{\H,\|},1/\delta_{m}]$.  The angular momentum $\bJ$ and the angle $\beta$ between $\bJ$ and $\Bv$ obtained from the LEs are employed to compute the net degree of alignment as follows: 
\bea
 R\equiv \sum_{n=0}^{N_{\rm step}-1} \frac{G_{X}(\cos^{2}\theta)G_{J}(\cos^{2}\beta_{n})}{N_{\rm step}},
 \ena
where $G_{X}$ can be replaced by $q_{X} (J_{n})=\int_{0}^{\pi}G_{X}(\cos^{2}\theta)f_{\rm LTE}(\theta, J_{n})d\theta$ with $f_{\rm LTE}(\theta,J_{n})\propto \exp(-J_{n}^{2}\left[1+(I_{\|}/I_{\perp}-1)\sin^{2}\theta\right]/2I_{\|}k_{B}T_{d})\sin\theta$ in the case of fast internal relaxation (see e.g, RL99). As usual, the initial grain angular momentum is assumed to have random orientation in the space and magnitude $J=I_{\|}\omega_{T}$ (i.e., $j=1$).

For magnetic grains, $h$ is determined essentially by two timescales $\tau_{\rm m}$ and $\tau_{\gas}$ (or a single parameter $\delta_{m}$), and the value of $\delta_{m}$ can be huge (i.e., $\delta_{m}\sim 10^{4}$). In this case, a tiny timestep $h \ll \delta_{\rm m}^{-1}$ is needed to achieve sufficient statistics, which requires a huge number of time steps (e.g., $N_{T}\sim 10^{8}-10^{9}$) and large computing time. In the following, a fixed integration time $T=10^{3}\tau_{\gas}$ is chosen, which ensures that $T$ is much larger than the longest dynamical timescale to provide good statistical calculations of the degrees of grain alignment.  We adopt the typical CNM phase of the interstellar medium for calculations.

\subsection{Big silicate grains with ferromagnetic inclusions}

We first quantify the magnetic alignment for thermally rotating big silicate grains incorporated with iron clusters, which give rise to superparamagnetic and ferromagnetic materials. Several values of grain size $a=0.05-0.3\mum$ and a typical dust temperature $T_{d}=20$K are considered. For each value of $a$, we vary $N_{\cl}$ from $1-10^{3}$ with volume filling factor $\phi_{\rm sp}\approx 0.03$. A variety of interaction processes is considered, although the dominant interaction for the $a>0.1\mum$ grains is gas-grain collisions (see Figure \ref{fig:FG}).

\begin{figure*}
\includegraphics[width=0.45\textwidth]{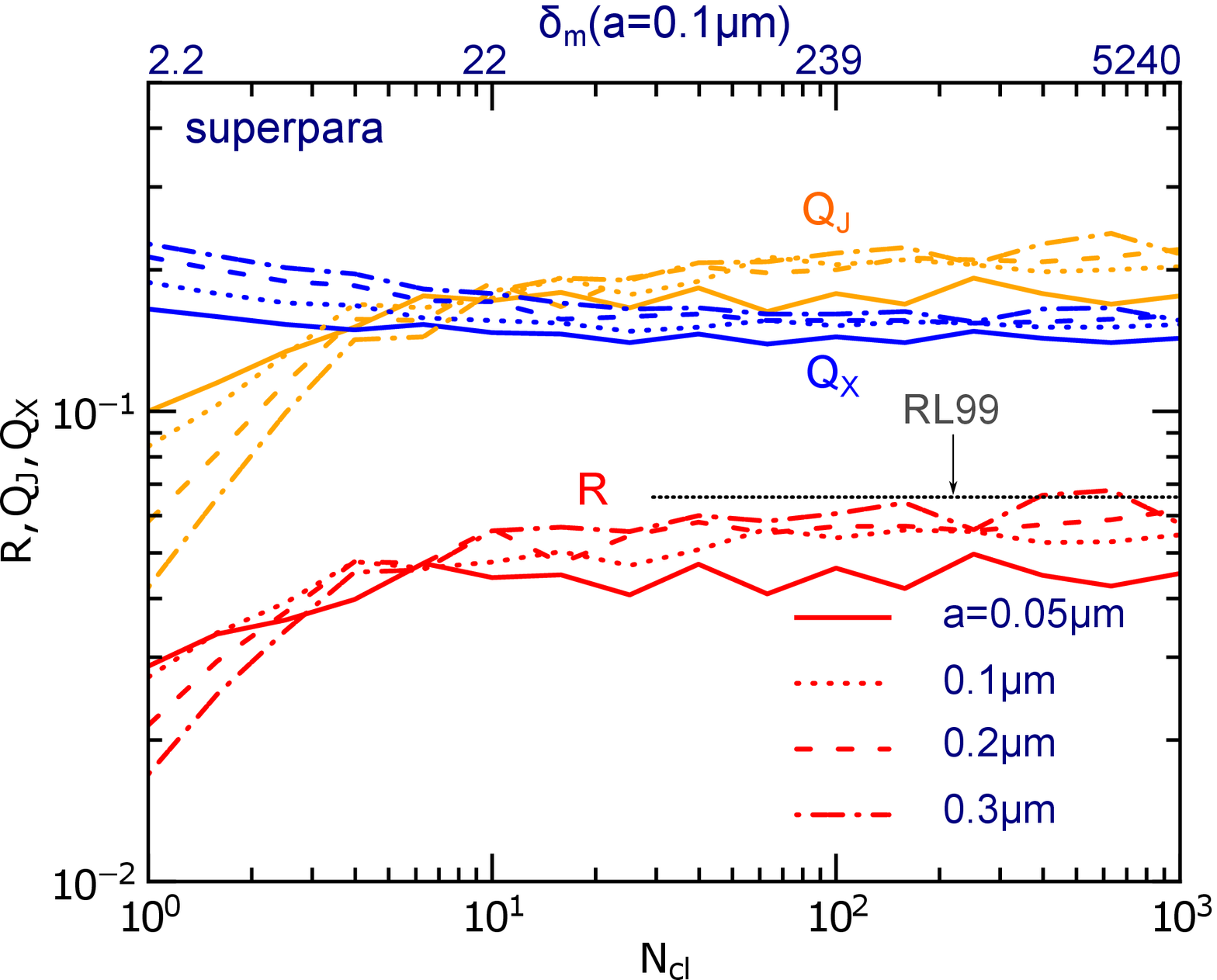}
\includegraphics[width=0.45\textwidth]{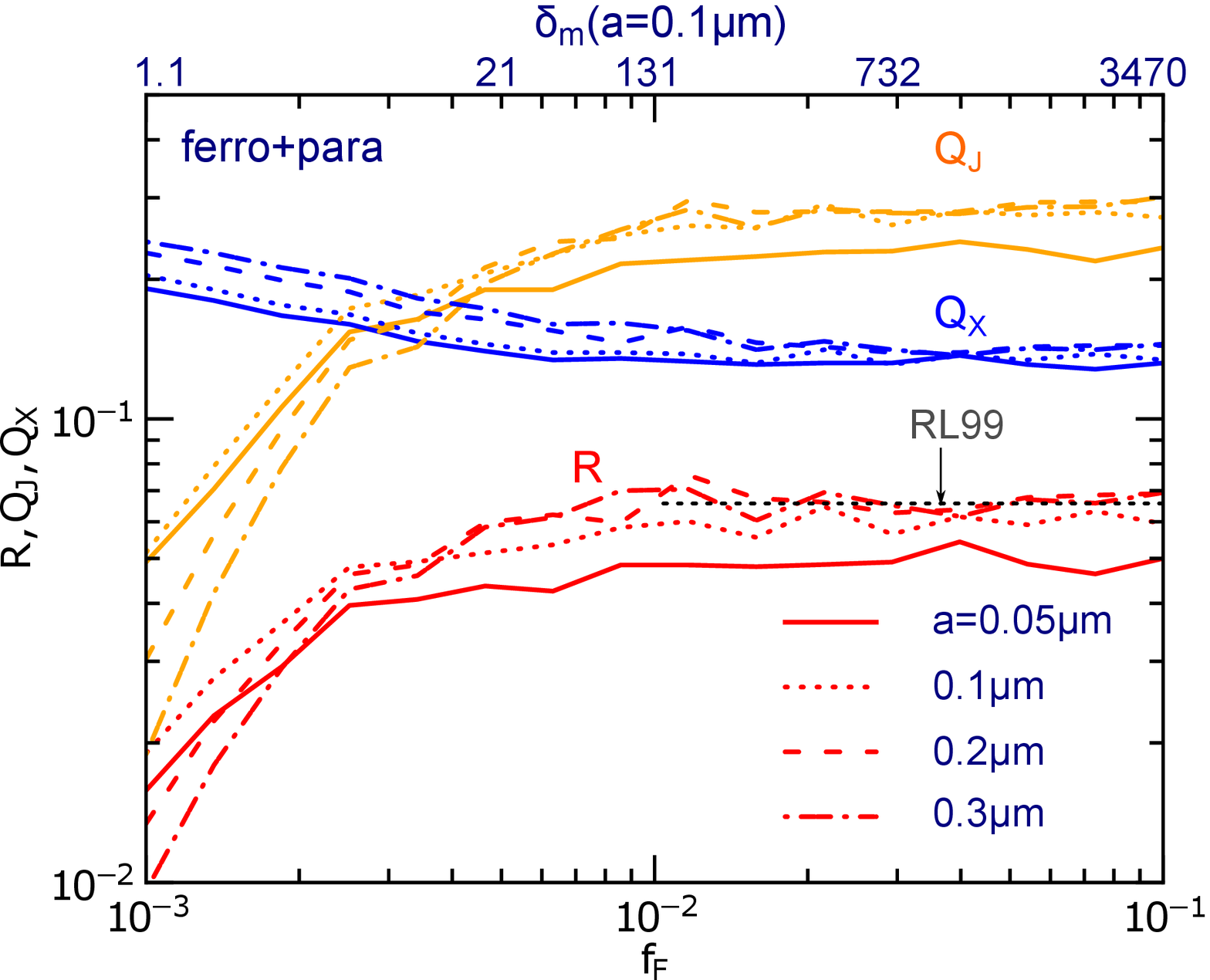}
\caption{Left panel: degrees of alignment of silicate grains with superparamagnetic inclusions of different $N_{\cl}$. Right panel: same as the left one, but for grains with ferro-paramagnetic interactions and for the different values of $f_{F}$. Four values of the grain size $a=0.05, 0.1, 0.2$, and $0.3\mum$ and grain temperature $T_{d}=20$K are considered. Dotted lines show the value of $R$ computed by RL99 for $\delta_{m}=100$ at the same environment conditions.}
\label{fig:Rali_superin}
\end{figure*}

Figure \ref{fig:Rali_superin} (left panel) shows the obtained values of $Q_{X}, Q_{J}$ and $R$ as functions of $N_{\cl}$ for superparamagnetic grains. The value of $Q_{J}$ rapidly increases with increasing $N_{\cl}$ ($\delta_{m}$) and slightly varies for $N_{\cl}> 10$ ($\delta_{m}>20$). $Q_{X}$ tends to decrease with increasing $N_{\cl}$ ($\delta_{m}$) as a result of faster magnetic dissipation. The value of $R$ rapidly increases with $N_{\cl}$ first and then becomes nearly independent on $N_{\cl}$ for $N_{\cl}>10$ (or $\delta_{m}> 20$), which is referred to as {\it alignment saturation regime}. Moreover, the alignment saturation occurs with $Q_{J}\le 0.4$ and $R\le 0.07$ for the assumed environment conditions. Bigger grains tend to have higher degree of alignment saturation due to their lower rotational damping (see Figure \ref{fig:FG}). 

Figure \ref{fig:Rali_superin} (right panel) shows $Q_{X}, Q_{J}$ and $R$ as functions of $f_{F}$--the fraction of the total atoms in the grains present in iron clusters (see Equation \ref{eq:K_ferropara}) when the effect of ferromagnetism on the paramagnetic grain is accounted for. The value of $Q_{J}$ tends to increase with increasing $f_{F}$, whereas $Q_{X}$ tends to decrease with increasing $f_{F}$ ($\delta_{m}$) as seen with superparamagnetic grains. The value of $R$ rapidly increase with $f_{Fe}$ first up to $f_{F}=0.01$ ($\delta_{m}\sim 130$) and reach alignment saturation with $R\le 0.07$ for $f_{F}>0.01$. 

{The alignment saturation observed for big grains with ferromagnetic inclusions is resulting from the detailed balance of magnetic dissipation and fluctuations when $\delta_{m}\gg 1$. In this saturation regime, the alignment becomes independent of the magnetic relaxation and only depends on other processes acting in the parallel direction to $\bJ$. We also found that, due to the additional IR damping included, our saturation levels are slightly lower than the semi-analytical results obtained in RL99 for the same conditions ($s=0.5, T_{d}/T_{\gas}=0.2$), as shown by dotted line in Figure \ref{fig:Rali_superin}.}

\subsection{Free-flying magnetic particles}

In the following, we are going to compute the degree of alignment for single-domain ferromagnetic particles with susceptibility given by Equation (\ref{eq:chi_cd}) for (\ref{eq:tau_DG}) for the range of grain size $a=0.001-0.1\mum$. %Note that grains larger than $a\sim 0.02\mum$ may be multiple-domain.

\subsubsection{Alignment subject to neutral-grain collisions}
\begin{figure}
\centering
\includegraphics[width=0.45\textwidth]{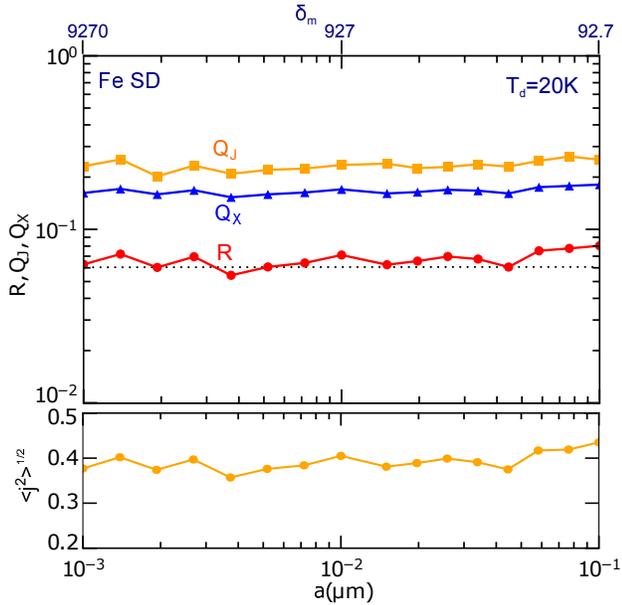}
\caption{Degrees of grain alignment of single-domain (SD) ferromagnetic particles as functions of grain size when only neutral-grain collisions and magnetic relaxation are considered. Lower panel shows the value of $\langle j^{2}\rangle^{1/2}$ in normalized units vs. $a$. The grain temperature $T_{d}=20$K is assumed. The horizontal dotted line marks the approximate level of alignment saturation.}
\label{fig:Rali_iron}
\end{figure}

We first consider the case in which grain rotational dynamics is purely induced by neutral-grain collisions and magnetic relaxation that follow detailed balance. The obtained results for $T_{d}=20$K and the DL99 susceptibility form are shown in Figure \ref{fig:Rali_iron}. The degrees of alignment appear to decrease slightly with decreasing the grain size from $a=0.1-0.001\mum$ or increasing $\delta_{m}$ from $\delta_{m}= 92.7-9270$ (upper panel). The similar trend is seen for the variation of $\langle j^{2}\rangle^{1/2}$ (lower panel). This {\it alignment saturation} is reminiscent of what seen in the case of big grains with ferromagnetic inclusions of $\delta_{m}\gg 1$. 

%The results for Fe$_3$O$_4$ and Fe$_2$O$_3$ structures are almost similar because they all have $\delta_{m}\gg 1$.

\subsubsection{Alignment in the presence of various interaction processes}
Now we take into account all interaction processes, including IR emission and ion collisions that do not follow detailed balance. Three values of $T_{d}=20, 30$ and $40$ K are considered. 

\begin{figure*}
\centering
\includegraphics[width=0.9\textwidth]{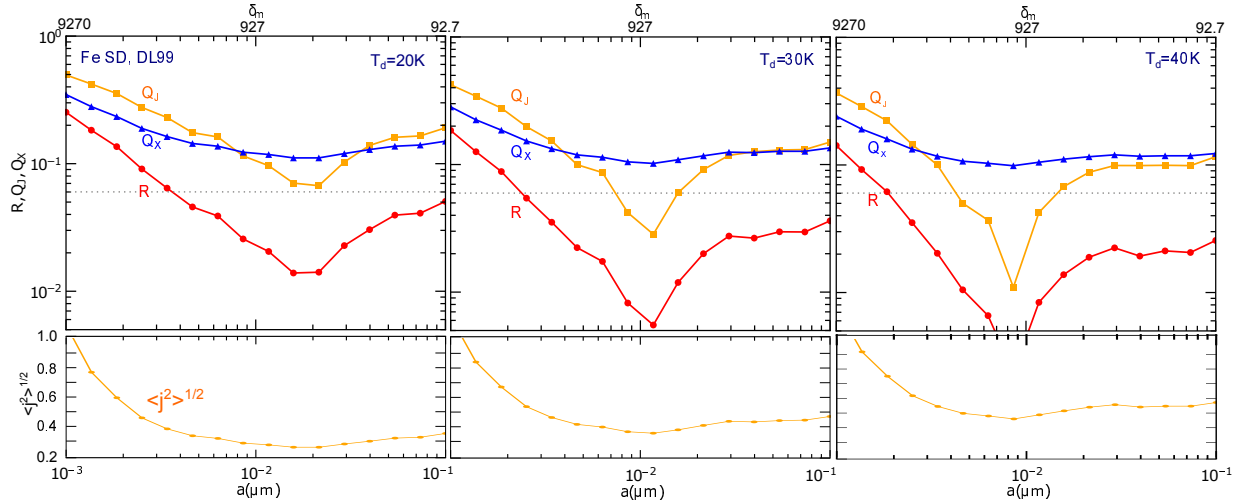}
\caption{Same as Figure \ref{fig:Rali_iron} but for the case ion collisions and IR emission are included. Three values of $T_{d}$ are considered. Horizontal dotted lines describe the approximate alignment saturation of $R$ from Figure \ref{fig:Rali_iron}. Strong damping by IR emission results in the minimum alignment at $a\sim 0.01\mum$, and the alignment of ultrasmall particles is enhanced by ion collisions. }
\label{fig:Rali_iron2}
\end{figure*}

Figure \ref{fig:Rali_iron2} show the results computed with the magnetic susceptibility from DL99. First, the degree of alignment does not exhibit alignment saturation but varies significantly with $a$ despite $\delta_{m}\gg 1$ (i.e., being in the saturation regime). When $a$ is decreased from $a=0.1\mum$, $Q_{J}$ and $R$ decrease with decreasing $a$ until their minimum around $a\sim 0.01\mum$ and reverse their trend for smaller $a$. In particular, the degree of alignment varies with $a$ in a similar trend as $\langle j^{2}\rangle^{1/2}$ (see the lower part of each panel). Second, higher $T_{d}$ result in lower degrees of alignment but higher value of $\langle j^{2}\rangle^{1/2}$ due to stronger thermal fluctuations (\citealt{1994MNRAS.268..713L}; \citealt{1997ApJ...484..230L}) (see lower panels). Moreover, the location of the minimum tends to shift to smaller grain sizes for higher $T_{d}$.

\begin{figure*}
\centering
\includegraphics[width=0.9\textwidth]{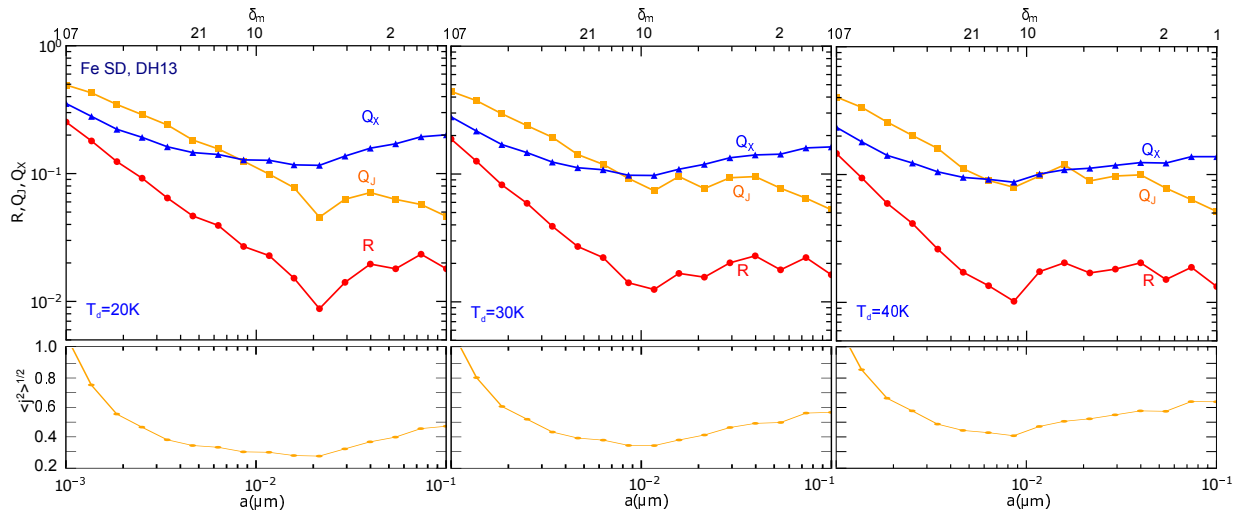}
\caption{Same as Figure \ref{fig:Rali_iron2} but the susceptibility of single-domain (SD) iron is taken from DH13 with $\alpha_{G}=0.2$. The degree of alignment increases rapidly with $a$ for ultrasmall grains, as in Figure \ref{fig:Rali_iron2}.}
\label{fig:Rali_iron2_DH}
\end{figure*}

Figure \ref{fig:Rali_iron2_DH} present the degrees of alignment computed with the magnetic susceptibility from DH13 (see Equation \ref{eq:chi2_DH}). The results exhibit remarkable similarity with those obtained with the DL99 susceptibility (see Figure \ref{fig:Rali_iron2}), in which the alignment minimum occurs at $a\sim 0.01\mum$, and the alignment increases rapidly with decreasing $a$ from that position. The obvious difference in alignment occurs for $a>0.01\mum$ where the magnetic relaxation rate is moderate of $\delta_{m}< 20$ (i.e., not yet in the alignment saturation regime) in which the value of $Q_{J}$ increases with increasing $\delta_{m}$.

{The disappearance of alignment saturation for ferromagnetic nanoparticles with $\delta_{m}\gg 1$ in the presence of ion-grain collisions and IR emission can be understood as follows. In the case of extreme magnetic dissipation ($\delta_{m}\gg 1$), the torques perpendicular to $\bJ$ are essentially the magnetic torque, while the torques parallel to $\bJ$ are determined by a variety of other interaction processes including imbalanced ones. The latter determine how fast the grain is spinning statistically, i.e., the value of $\langle j^{2}\rangle^{1/2}$ that determines the ultimate degree of magnetic alignment. Therefore, when the excitation by ion collisions increases and/or the damping by IR emission decreases (see Figure \ref{fig:FG}), it results in an increase in $\langle j^{2}\rangle^{1/2}$ and then the degree of alignment of ultrasmall grains of $a<0.005\mum$ (see Figure \ref{fig:Rali_iron2} and \ref{fig:Rali_iron2_DH}). The increase of IR damping results in the decrease of alignment of small particles of $a\sim 0.01-0.05\mum$. }

\section{Polarization of magnetic dust emission}\label{sec:mdepol}
\begin{figure*}
\centering
\includegraphics[width=0.8\textwidth]{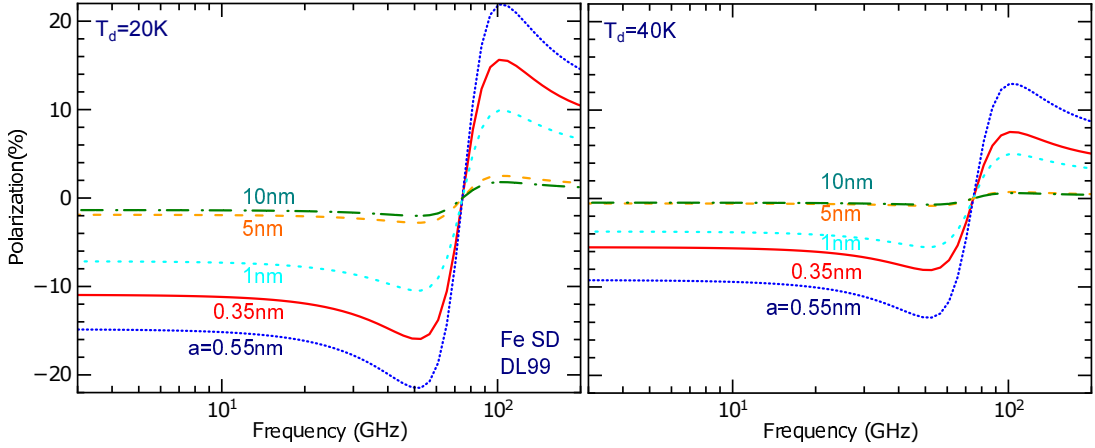}
\caption{Polarization degree of MDE by free-flying magnetic particles of different radii calculated with our numerical degrees of alignment for $\gamma_{B}=0$ for two values of $T_{d}$. Results for the oblate spheroid with $a_{1}:a_{2}:a_{3}=1:2:2$ are presented. The results obtained with the susceptibility from DL99, and the reverse of polarization sign occurs at $\nu\sim 75$GHz.}
\label{fig:MDE_pol}
\end{figure*}

\begin{figure*}
\centering
\includegraphics[width=0.8\textwidth]{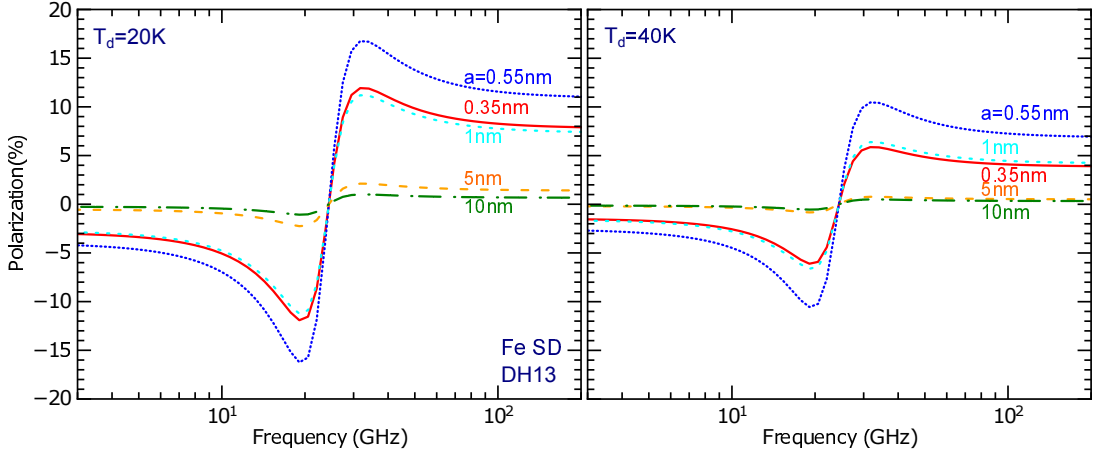}
\caption{Same as Figure \ref{fig:MDE_pol} but for the results computed for the susceptibility from DH13. The reverse of polarization sign shifts to the lower frequency at $\nu \sim 24$GHz due to the lower susceptibility compared to DL99.}
\label{fig:MDE_pol_DH}
\end{figure*}

{The polarization of MDE from free-flying magnetic nanoparticles is calculated as the following:
\bea
p=\frac{I_{x}-I_{y}}{I_{x}+I_{y}},
\ena
where $I_{x}$ and $I_{y}$ are the radiation intensity with the electric field $\bE$ parallel to $\xhat$ and $\yhat$ in the plane of the sky  (see Appendix \ref{apdx:pol} for more details). 

In the presence of the imperfect alignment, $I_{x}$ and $I_{y}$ are given by
\bea
I_{x}-I_{y}& =&\int_{a_{\min}}^{a_{\max}} C_{\pol}R\cos^{2}\gamma_{B} B_{\nu}(T_{d})\frac{dn_{m}}{da} da,\\
I_{x}+I_{y} & =&\int_{a_{\min}}^{a_{\max}} \left[2C_{\rm avg} +RC_{\pol} \cos^{2}\gamma_{B}/3(2/\cos^{2}\gamma_{B}-3)\right]\nonumber\\
&&\times B_{\nu}(T_{d})\frac{dn_{m}}{da} da,
\ena
where $C_{\pol}=C_{\perp}-C_{\|}$, $C_{\rm avg}=(2C_{\perp} + C_{\|})/3$, $dn_{m}/da$ is the size distribution of magnetic particles, $B_{\nu}(T_{d})$ is the Planck function, and $\gamma_{B}$ is the angle between the magnetic field and the plane of the sky (see \citealt{1985ApJ...290..211L}; RL99).

\subsection{Single-size magnetic nanoparticles}
First, we assume that magnetic particles have the size distribution $n_{\H}^{-1}dn_{m}(a')/da=Cn_{\H}^{-1}n(a)\delta(a'-a)$ (i.e., grains of single size $a$). The relative number density of magnetic particles per H is $n(a)/n_{\H}= 3V_{Y}/(4\pi a^{3})$, where $V_{Y}$ is the grain volume per H of material Y. Following DL99,  $V_{Y} = 0.3f_{Y}(4/\rho_{Y}z_{Y})V_{sil,0}$, where $f_{Y}$ and $z_{Y}$ are the fraction of total Fe incorporated and the fraction of Fe mass of material $Y$, respectively. Here $V_{sil,0}=2.5\times 10^{-27}\cm^{3}/\H$ is the grain volume per H for a typical material $Mg_{1.1}Fe_{0.9}SiO_{4}$ that accommodates the entire budgets of Mg, Si, and Fe (see DL99). 

For metallic iron with $\rho_{Y}=7.86\g\cm^{-3}$ and $z_{Y}=1$, it follows that $V_{gr}\simeq 0.15V_{sil,0}f_{Y}\simeq 3.84\times 10^{-28}f_{Y} \cm^{3}/\H$. Similarly, we obtain $V_{Y}\simeq 0.32V_{sil,0}f_{Y}\simeq 8.01\times 10^{-28}f_{Y} \cm^{3}/\H$ for $Fe_{3}O_{4}$, and $V_{Y}\simeq 8.78\times 10^{-28}f_{Y}\cm^{3}/\H$ for $\gamma Fe_{2}O_{3}$ with $\rho=5.2\g\cm^{-3},~z_{Y}=0.72$ and $4.88\g\cm^{-3},~z_{Y}=0.7$, respectively. 

Figure \ref{fig:MDE_pol} shows the polarization spectra of MDE computed for an oblate spheroid of single sizes using the DL99 susceptibility. First, the polarization reverses its sign from negative to positive at $\nu\sim 75$ GHz where electric dipole emission becomes dominant. Moreover, MDE from smaller grains has higher degree of polarization, as expected from the computed degrees of alignment (see Figure \ref{fig:Rali_superin}). In particular, if iron particles are larger than $1$ nm, then, the polarization of its MDE is below $5\%$. 

Figure \ref{fig:MDE_pol_DH} shows the polarization spectra obtained using the susceptibility form from DH13. The polarization is essentially lower than shown in Figure \ref{fig:MDE_pol_sizedist} due to its lower degree of alignment. In particular, due to the lower susceptibility, the reverse in polarization sign occurs at lower frequency than in the case of the DL99 susceptibility (i.e. electric dipole emission dominate magnetic emission at lower frequency).
}

\subsection{Size distribution of magnetic particles}
Now let us assume that all Fe abundance is present in the form of small clusters of radius from $a_{\min}-a_{\max}$. At low temperatures, even very small iron clusters are ferromagnetic (i.e., having spontaneous magnetization), so we can take $a_{\min}=0.35$nm or $N_{\cl}\sim 20$ \citep{1994Sci...265.1682B}. The upper size is chosen as the critical size of single-domain, i.e., $a_{\max}=20$nm. Because there is no clue about any specific size distributions, we assume a typical power law $n_{\H}^{-1}dn_{m}/da = Ca^{-3.5}$ (\citealt{Mathis:1977p3072}), where $C$ is a normalization constant determined by the grain volume per H:
\bea
\int_{a_{\min}}^{a_{\max}} \frac{4\pi a^{3}}{3}\frac{n_{\H}^{-1}dn_{m}}{da} da=\frac{8\pi C}{3}\left(a_{\max}^{1/2}-a_{\min}^{1/2} \right)=V_{gr}.
\ena

\begin{figure*}
\centering
\includegraphics[width=0.45\textwidth]{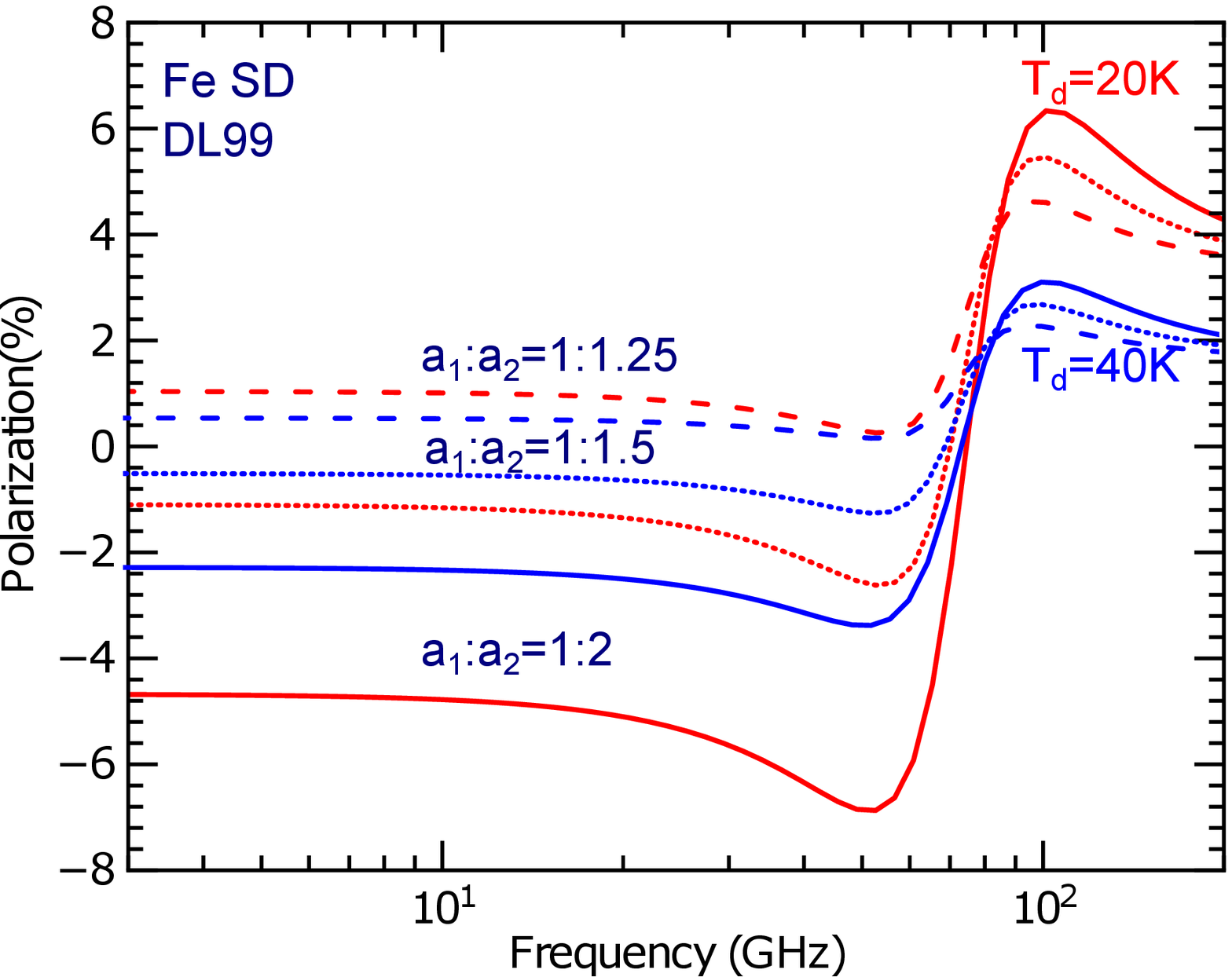}
\includegraphics[width=0.45\textwidth]{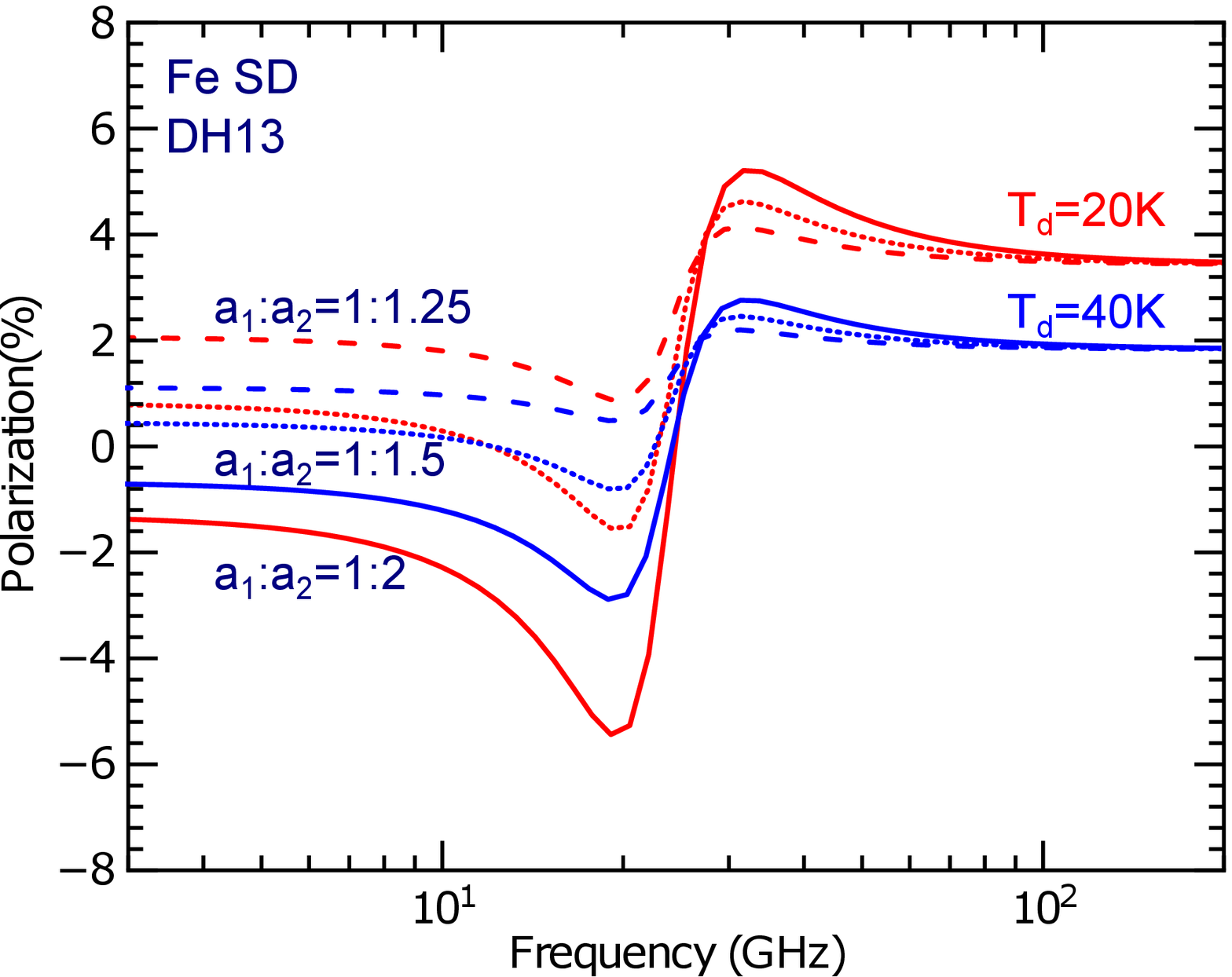}
\caption{Same as Figure \ref{fig:MDE_pol}, but the grain size distribution from $a=0.35-20$nm is considered. Left and right panel show the results obtained with the susceptibility from DL99 and DH13, respectively. Three different oblate spheroids are considered.}
\label{fig:MDE_pol_sizedist}
\end{figure*}

The polarization spectra are shown in Figure \ref{fig:MDE_pol_sizedist} two cases of the DL99 (left panel) and DH13 (right panel) susceptibility. We can see that the polarization of integrated MDE is rather low (e.g, less than $5\%$ for $T_{d}=40$K), which is substantially lower than the MDE polarization from nanoparticles of single radius of $a<1$nm. This feature is due to the fact that the intensity of MDE depends on the grain volume for which the largest particles dominate the MDE. However, those larger particles are found to have the lower degree of alignment due to stronger IR damping (see Figure \ref{fig:Rali_iron2}). The polarization using DH13 susceptibility is slightly lower due to lower susceptibility.

\section{Spinning dust emission by free-flying magnetic particles}\label{sec:spinem}
Spinning magnetic particles with permanent magnetic moments emit microwave emission by the same mechanism as spinning electric dipole emission (\citealt{Erickson:1957p4806}). The spinning emissivity can be computed by the same model of HDL10, but the electric dipole moment $\mu_{\ed}$ is replaced by the magnetic dipole $\mu_{\md}$. We consider oblate spheroidal grains with axial ratio $r=2$ for a typical model.

The polarized emissivity and unpolarized emissivity of spinning dust emission are calculated as follows:
\bea
q_{\nu}&=&\int_{a_{\min}}^{a_{\max}} Q_{J}(a)\cos^{2}\gamma_{B} j_{\nu}(a) \frac{n_{\H}^{-1}dn_{m}}{da} da ,\\
j_{\nu}&=&\int_{a_{\min}}^{a_{\max}} j_{\nu}(a) \frac{n_{\H}^{-1}dn}{da}da, 
\ena
where $j_{\nu}(a)$ is the spinning dust emissivity at frequency $\nu$ from a grain of size $a$, and $a_{\min}=0.35$nm is assumed. As in the previous section, we consider two cases of single-size particles and those following the MRN distribution. The polarization spectrum of spinning dust emission is $p(\nu)=q_{\nu}/j_{\nu}$. 

\begin{figure*}
\centering
\includegraphics[width=0.8\textwidth]{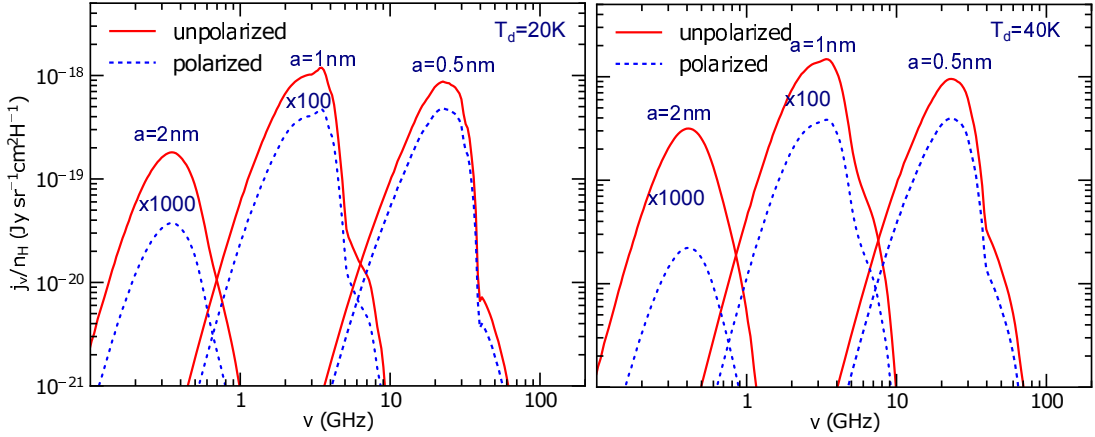}
\caption{Spinning dust emissivity from single-size ferromagnetic particles of single size for $T_{d}=20$K (left) and $40$K (right). Results for $a=1$nm and $2$nm are multiplied by 100 and 1000. Higher temperature slightly increases the emissivity but decreases the polarization.}
\label{fig:jnu_nano}
\end{figure*}

Figure \ref{fig:jnu_nano} shows the unpolarized emissivity and polarized emissivity for the single-size case with $a=0.5, 1$ and 2 nm. Smallest Fe particles emit considerable spinning emissivity, but it is still one order of magnitude lower than that from spinning PAHs (see, e.g.,  Figure \ref{fig:jnu_PAH}). We also computed the spinning emission from ferrimagnetic particles ($\gamma$Fe$_{2}$O$_{3}$ and Fe$_{3}$O$_{4}$) and found that their emission is much lower than spinning metallic iron, which is expected due to their much smaller $M_{s}$. Notably, the polarization degree of spinning emission from iron particles is rather high, up to $40-50\%$ for $T_{d}=20-40$K. This is a direct result of the efficient external alignment $Q_{J}$ of tiny iron particles as shown in Figure \ref{fig:Rali_iron}.

\begin{figure*}
\centering
\includegraphics[width=0.8\textwidth]{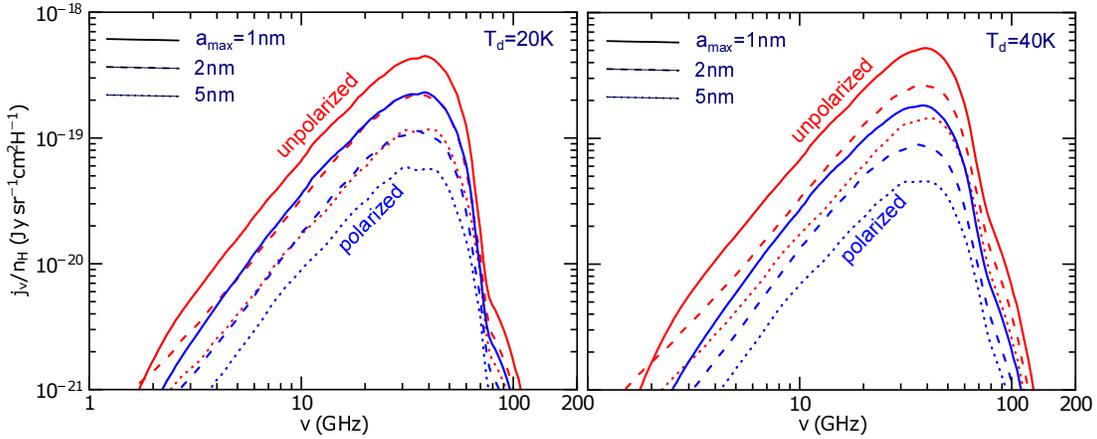}
\caption{Same as Figure \ref{fig:jnu_nano}, but for the case magnetic particles follow the MRN distribution with $a_{\max}=1, 2$ and 5 nm.}
\label{fig:jnu_sizedist}
\end{figure*}

\begin{figure*}
\centering
\includegraphics[width=0.8\textwidth]{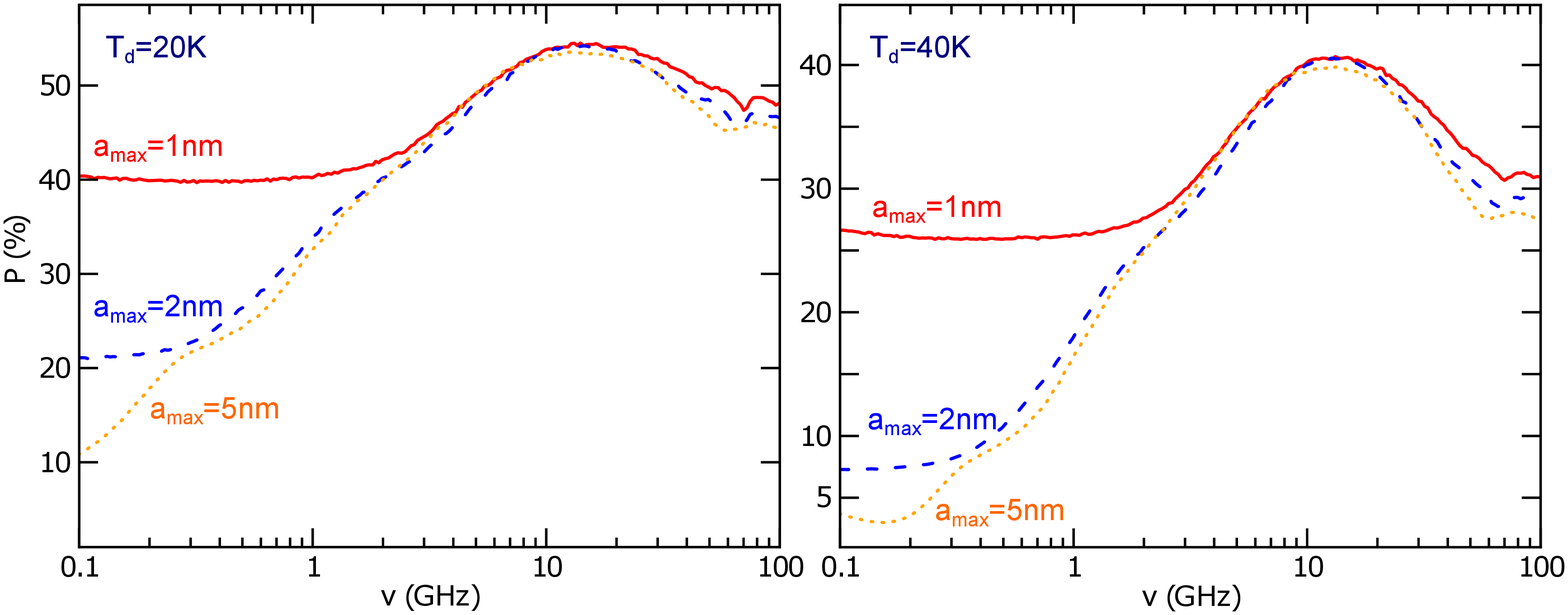}
\caption{Polarization spectra of spinning dust emission from magnetic nanoparticles with the MRN size distribution and different values of $a_{\max}$. High polarization up to $50\%$ is observed.}
\label{fig:pol_sizedist}
\end{figure*}

Figure \ref{fig:jnu_sizedist} shows the emissivity for the case of the MRN distribution with $a_{\max}=1, 2$ and 5 nm. The spinning emission is strongest when all iron atoms are concentrated in the range $a \le 1$ nm. The peak emissivity is insensitive to $a_{\max}$, which is expected from the fact spinning emission is dominated by smallest grains having the fastest rotation and the contribution from larger grains is negligible (see Figure \ref{fig:jnu_nano}). The polarization of spinning emission from iron particles can reach $40-50\%$ for the magnetic field in the plane of the sky (see Figure \ref{fig:pol_sizedist}), which originate directly from high degree of alignment (i.e., $Q_{J}$) of ultrasmall iron particles (see Figure \ref{fig:Rali_iron2}).

\section{Discussion}\label{sec:discus}
\subsection{Magnetic dipole emission from magnetic nanoparticles}\label{sec:DL_DH}
{Magnetic dipole emission (MDE) was first suggested in DL99 as a potential source of AME for frequencies of $\nu\sim 10-60$ GHz. In this original model, the magnetic response of the ferromagnetic material to an applied oscillating magnetic field is treated using the Drude model with scalar susceptibility. Recently, DH13 revisited the treatment of magnetic response by using the Gilbert equation with tensor susceptibility, where the magnetization damping is characterized by a dimensionless Gilbert parameter $\alpha_{G}$. With a new form of magnetic susceptibility, DH13 found that MDE is important for frequencies of $\nu=70-300$ GHz.

Thus, more experimental and theoretical studies are undoubtedly necessary to better understand the susceptibility of magnetic material at high frequencies and to differentiate the emission spectrum of MDE predicted by the two models. The pressing issue now is to have quantitative, realistic predictions for the polarization of MDE for reliable detection of CMB B-mode signal. For this purpose, we have used the magnetic susceptibility from both DL99 and DH13 to compute the degree of alignment for ferromagnetic nanoparticles and obtained the MDE polarization spectra for both models. 

\subsection{Degree of alignment of thermally rotating magnetic grains}
The first study on magnetic alignment of thermally rotating grains that takes into account the effect of imperfect Barnett relaxation (\citealt{1979ApJ...231..404P}; \citealt{1997ApJ...484..230L}) was performed in \cite{1997MNRAS.288..609L} using the analytical method. RL99 used the numerical method based on the Langevin equation to compute the degree of alignment for both paramagnetic and superparamagnetic grains of different values of axial ratio $s$, $T_{d}/T_{\gas}$ and $\delta_{m}$. The RL99's study only considered the alignment of big grains subject to neutral-grain collisions and superparamagnetic relaxation of $\delta_{m}\le 100$. 

The present work first improved the numerical method in RL99 by implementing a second-order integrator for solving the Langevin equations, and extended calculations for a wider range of $\delta_{m}\sim 1-10^{4}$. We find that thermally rotating superparamagnetic grains are weakly aligned and achieve {\it alignment saturation} of $R<0.1$ for $\delta_{m}\gg 1$. We also have quantified the effect of ferromagnetic and paramagnetic interactions on the alignment of big silicate grains for a wide range of the fraction of iron in clusters $f_{F}=0.001-0.1$. Same as thermally rotating superparamagnetic grains, the degree of alignment is low and becomes saturated for $f_{F}>0.01$ or $\delta_{m}\ge 100$. One consequence of the alignment saturation for the grains with superparamagnetic/ferromagnetic inclusions is that the dust polarization becomes independent on the magnetic field strength (cf. HLM14) and depends mostly on the magnetic field direction.

Second, we have computed the degrees of alignment of single-domain ferromagnetic nanoparticles, accounting for various rotational damping and excitations by neutral-grain collisions, ion collisions, and IR emission, using the susceptibility form from both DL99 and DH13. It is noted that the ultrasmall iron particles can have very large value of $\delta_{m}\sim 10^{4}$, which requires significant computing time to obtain good statistics. {Interestingly, we found that the alignment saturation expected for $\delta_{m}\gg 1$ is destroyed in the presence of imbalanced interaction processes. Specifically, we identify that rotational excitation from ion collisions, which becomes more efficient for tiny grains of negatively charge, can enhance the alignment of very small iron grains ($a<0.005\mum$), whereas strong IR damping can significantly reduce the alignment of $a\sim 0.01-0.05\mum$ particles. This result indicates that even subtle excitation/damping by imbalanced processes (e.g., ion collision and IR emission) can have a great impact on the alignment of grains in the alignment saturation regime $\delta_{m}\gg 1$. Notably, we found no clear difference in the degree of alignment of very small iron grains between the susceptibility form from DH13 and DL99. This is due to the fact that both models yield $\delta_{m}\gg 1$ (i.e., saturation regime) such that the alignment efficiency becomes independent of the magnetic relaxation rate.} 

Lastly, the low degrees of alignment for thermally rotating, superparamagnetic grains in realistic environmental conditions (i.e., $T_{gas}$ not much larger than $T_{d}$) indicate that suprathermal rotation is required for producing efficient alignment, as required by observations (see latest reviews in \citealt{Andersson:2015bq}; \citealt{LAH15}). If magnetic nanoparticles are incorporated into big grains, then such grains have greatly enhanced magnetic susceptibility, either through superparamagnetic clusters or the ferro-paramagnetic interaction. As a result, the joint action of radiative torques and the enhanced magnetic dissipation would likely lead to efficient alignment \citep{Lazarian:2008fw}. Detailed study will be presented in our future paper.

\subsection{Polarization of magnetic dipole emission: free-fliers vs. magnetic inclusions}

Magnetic nanoparticles, whether being free-fliers or inclusions, produce slightly different MDE spectra (DL99; DH13), while the polarization of MDE from these two forms is usually believed to be distinct. For instance, assuming its perfect alignment with the interstellar magnetic field, DL99 predicted that free-flying iron spheroids produce high polarization level ($P\sim 10-30\%$ for $\nu < 30$ GHz), whereas randomly oriented ferromagnetic inclusions would produce low polarization level. In DH13, the MDE polarization is also expected to be large (i.e., $P$ can reach $\sim 30\%$) for free-flying magnetic nanoparticles and a lower level for big grains with magnetic inclusions. Yet, we have found that the polarization level of MDE from free-flying iron nanoparticles is rather small for both the DL99 and DH13 susceptibility forms.

Indeed, MDE by free-flying magnetic particles increases with the particle volume, so does its polarization level. For the size range of $a=0.35-20$nm where single-domain iron clusters are expected to exist in the ISM conditions, the largest particles dominate MDE and its polarization level. Meanwhile, we found that these large particles tend to have much lower degree of alignment than the smallest ones, with the minimal alignment at $a\sim 10$nm due to strong IR damping. Therefore, unless magnetic particles are ultrasmall of $a<1$nm for which the MDE polarization can reach 20$\%$ ($10\%$) for $T_{d}=20$K (40K), the polarization of MDE is predicted to be below $5\%$ at $T_{d}=40$K (see Figure \ref{fig:MDE_pol} and \ref{fig:MDE_pol_sizedist}). The polarization spectrum for the DH13 susceptibility reverses its sign at $\nu\sim 24$GHz, lower than $\nu\sim 75$GHz for the DL99 model, which is directly resulting from its lower magnetic susceptibility.

It is noted that the polarization of MDE from magnetic inclusions depends on the alignment of big silicate grains, which tends to vary with the local radiation intensity according to the radiative alignment theory (\citep{LAH15}). On the other hand, the MDE polarization from free-flying particles does not depend on the radiation field. Thus, it may be possible to search for the form of magnetic particles by studying the MDE polarization from the diffuse medium and dense clouds. If the variation of the MDE polarization is observed toward the denser regions, then, the MDE may be produced by magnetic inclusions instead of free-flying particles.

\subsection{Polarization of spinning dust emission from magnetic particles}
\subsubsection{Free-flying magnetic nanoparticles}
We have investigated spinning emission by spontaneous magnetization of magnetic particles and found that the largest emissivity is still one order of magnitude lower than spinning PAHs when all iron budget is present in single-size nanoparticles with $a\le 1nm$. One distinct feature of spinning emission from ferromagnetic particles is of its high degree of polarization, which can reach $40-50\%$ (see Figure \ref{fig:pol_sizedist}). Such a difference arises from the fact ferromagnetic nanoparticles have $\delta_{m}\gg 1$ and do not suffer the suppression of magnetic relaxation due to fast rotation (i.e., $\omega$ still lower than the spin-relaxation rate $1/\tau_{2}$). In addition, the rotational damping by spinning radiation that removes the grain angular momentum is very weak for magnetic particles due to its small dipole moment.

\subsubsection{Big grains with magnetic inclusions}
Big silicate grains are likely spinning suprathermally due to radiative torques (RATs; \citealt{1996ApJ...470..551D}; \citealt{2007MNRAS.378..910L}) and pinwheel torques (e.g., due to H$_{2}$ formation; \citealt{1979ApJ...231..404P}; \citealt{2009ApJ...695.1457H}). The existence of ferromagnetic inclusions in such suprathermally rotating big grains would induce rotational radiation. 

Let us assume that iron clusters are randomly oriented in a big nonmagnetic grain. The net magnetic moment of the grain is estimated to be
\bea
\mu_{\rm md}^{2} = \mathcal{N}_{\cl}m^{2}=3.5\times 10^{8}\phi_{\sp}N_{\cl}^{-1}a_{-5}^{3}m^{2},
\ena
where $m=N_{cl}p\mu_{B}$ is the magnetic moment of each cluster. 

The emission power by the grain rotating at $\omega$ is equal to
\bea
P &=& \frac{2\mu_{\rm md}^{2}\omega^{4}}{3c^{3}}\nonumber\\
&\simeq& 2.66\times 10^{-40}a_{-5}^{-7}\phi_{\sp}N_{\cl}\hat{T}_{\gas}^{2}(\omega/\omega_{T})^{4}{\erg\s^{-1}},\label{eq:Pmd}
\ena

For the MRN distribution, we can estimate the total emission intensity per H atom as follows:
\bea
S &=& \int_{a_{\min}}^{a_{\max}} P \frac{n_{\H}^{-1}dn}{da}da,\nonumber\\
&\simeq& 2.2\times 10^{-62}A_{\rm MRN} \left[a_{\max}^{1/2}-a_{\min}^{1/2} \right] \phi_{\sp}N_{\cl}\omega^{4},\nonumber\\
&\simeq&  1.2\times 10^{-44} \left[a_{\max,-5}^{1/2}-a_{\min,-5}^{1/2} \right] \phi_{\sp}N_{\cl}(\omega/\omega_{T,-5})^{4} \frac{\rm Jy \cm^{2}Hz}{\H},
\ena
where $A_{\rm MRN}=10^{-25.16}\cm^{-2.5}$, and $\omega_{T,-5}=\omega_{T}(a=10^{-5}\cm)$. 

If RATs are the main spin-up processes, we have $\omega \sim 10^{2}\omega_{T}\mathcal{G}$ \citep{2009ApJ...695.1457H} for $a\sim 0.1\mum$ where $\mathcal{G}$ is the ratio of the radiation energy relative to the average radiation density in the solar neighborhood. Thus, one has $S\simeq 2.5\times 10^{-36}\mathcal{G}^{4} {\rm Jy}\cm^{2}Hz/\H$, or $j_{\nu}=S/\nu \sim 10^{-43}\mathcal{G}^{3}{\rm Jy}\cm^{2}/\H$ where $\nu=\omega/2\pi$. The emissivity is clearly negligible compared to that from spinning PAHs in the ISM with $\mathcal{G}=1$. In the star forming regions with strong radiation (i.e., $\mathcal{G}\gg 1$), the spinning emission can be in the radio with $\omega\sim 10$ GHz, but it appears that the intensity of spinning radiation is unimportant.

{Our above estimates are for the rotation around the grain symmetry axis. In fact, due to incomplete internal relaxation (\citealt{1994MNRAS.268..713L}; \citealt{1997ApJ...484..230L}) grains may not rotate around the axis of maximal inertia, resulting in wobbling motion. As a result, the emission by an isolated, spinning triaxial grain occurs at many modes, which allows more energy to be extracted from the rotational energy (HDL10; \citealt{2011ApJ...741...87H}, hereafter HLD11). As a result, the total emissivity can be increased by a factor of 2, and peak frequency is increased by a factor 1.5 (HDL10). Moreover, the effect of transient spin-up by ion collisions and turbulence can also enhance the spinning dust emission. It seems that even including all these effects, spinning emission from big grains of magnetic inclusions is still negligible.}

\subsection{Polarization of spinning dust emission from PAHs}

During the last few years, we have witnessed significant progress in improving the dynamical models of spinning dust emission (HDL10; HLD11; \citealt{Silsbee:2011p5567}). Meanwhile, the polarization of spinning dust emission, which is essential for reliable measurements of CMB B-mode signal, is still uncertain. Nevertheless, its polarization is certainly determined by the alignment of PAHs, which is believed due to resonance paramagnetic relaxation (\citealt{2000ApJ...536L..15L}; HLM14). Using inversion technique combined with theoretical calculations, \cite{2013ApJ...779..152H} found a level of $1.6\%$ polarization for the line of sight toward a star (HD 197770) with the 2175$\AA$ polarization bump.

In Figure \ref{fig:pol_spinPAH}, we present the theoretical polarization spectra of spinning emission from PAHs in the CNM, computed for the different magnetic field strength with direction in the plane of the sky ($\gamma_{B}=0$) at $T_{d}=40$ and $60$K. The value of $f_{p}=0.01$ is assumed.  For the typical value of $B\sim 10\mu$ G, the polarization is less than $5\%$ for $\nu>20$ GHz. Colder spinning PAHs tend to generate stronger polarized emission due to weaker internal thermal fluctuations. Particularly, the polarization by spinning PAHs is lower than that by spinning Fe nanoparticles because the latter has much higher $\delta_{m}$ and does not suffer magnetic suppression due to the rapid rotation because the spin-spin relaxation rate still exceeds the rotation rate. 

The theoretical predictions for the polarization of spinning PAH emission appear to be consistent with the available observational data. For instance, several observational studies (\citealt{2006ApJ...645L.141B}; \citealt{2009ApJ...697.1187M}; \citealt{2011MNRAS.418L..35D}; \citealt{Battistelli:2015dt}) have found the upper limits for the AME polarization to be between $1\%$ and $5\%$. An upper limit of $1\%$ is also reported in various studies (see \citealt{LopezCaraballo:2011p508}; \citealt{RubinoMartin:2012ji}). Latest results by Planck (\citealt{2015arXiv150606660P}) show the AME polarization to be $0.6\pm 0.5\%$. 

\begin{figure*}
\centering
\includegraphics[width=0.8\textwidth]{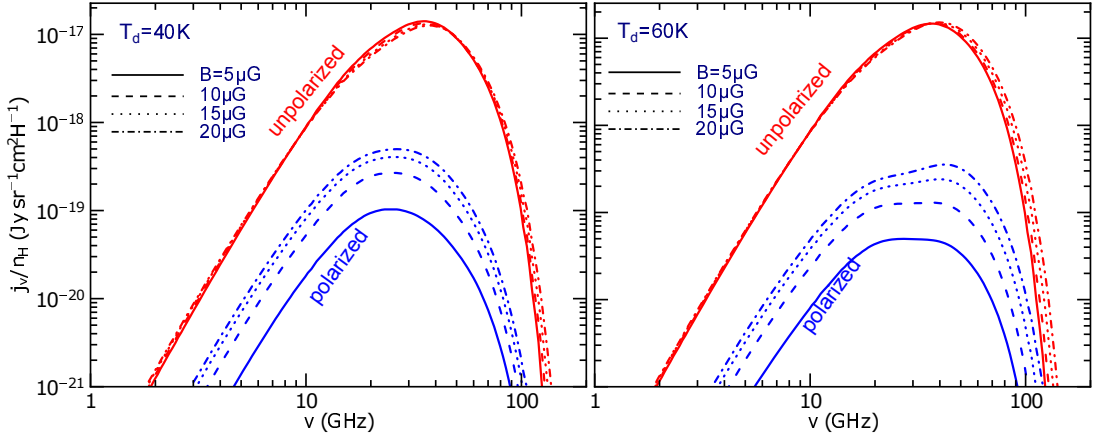}
\caption{Spinning dipole emission by electric dipole from PAHs computed for different magnetic field strengths. Two values of $T_{d}$ are considered. Magnetic fluctuations slightly increase the unpolarized spinning emissivity, but magnetic relaxation significantly increases the degree of alignment and then polarized emissivity.}
\label{fig:jnu_PAH}
\end{figure*}

\begin{figure*}
\centering
\includegraphics[width=0.8\textwidth]{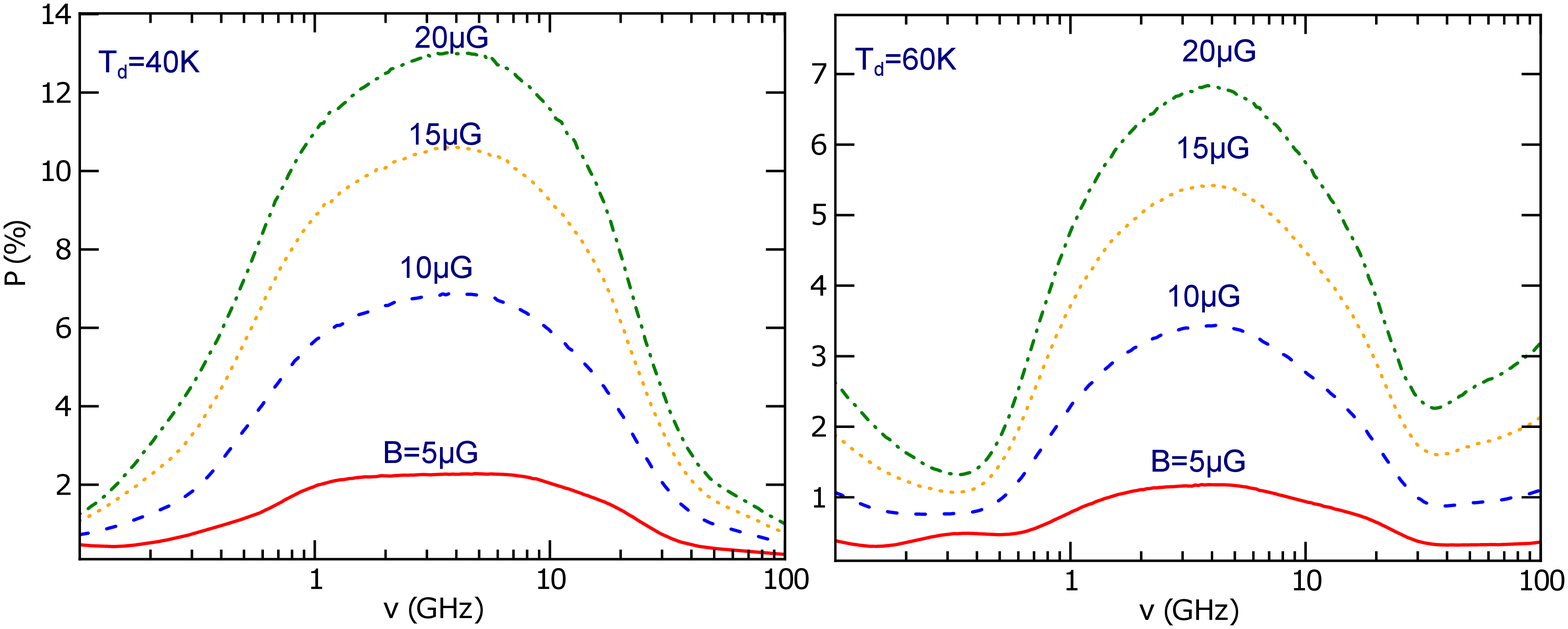}
\caption{Same as Figure \ref{fig:jnu_PAH}, but for the polarization spectra. Polarization degree is increased with the magnetic field strength for $B=5-20\mu$G.}
\label{fig:pol_spinPAH}
\end{figure*}

\subsection{On Mysterious Origin of AME}
Numerous observations have revealed that AME is most likely produced by rapidly spinning tiny dust grains (mostly PAHs) (see e.g., \citealt{PlanckCollaboration:2011hw}; \citealt{2015arXiv150606660P}). Especially, the low polarization level predicted for spinning dust emission from PAHs is in good agreement with the observational data (see the previous section).

\cite{Hensley:2015wca} (hereafter HDM) found a weak correlation of AME with PAH abundance using all-sky AME map, which reveals that AME might not originate from spinning PAHs. The paper suggests reconsideration of other origins for AME. {If the spinning PAH theory as it is formulated in DL98 and improved in HDL10, indeed, has problems explaining the observational data as claimed in HDM, we suggest several alternatives:
(a) Physics of PAH is more complex than it was assumed by DL98, e.g. the dipole moment changes with the environment in a way to offset the predicted correlations. (b) The emission is still from spinning dust, but the dust of not PAH nature. Iron and silicate nanoparticles are the primary candidate. (c) The role of magnetic dust emission should be reconsidered. (d) The AME does not originate from dust, but has a completely different nature.

The physics that can change PAHs and then its spinning emission to satisfy the analysis of HDM is unclear. However, there has not been enough studies of interstellar PAHs to dismiss this opportunity. This leaves the case (a) somewhat in limbo.

Regarding the issue (b), in the present paper, we have quantified spinning emission from iron nanoparticles and found that its rotational emission is maximum within the AME frequency range but has low emissivity of $j_{\nu}/n_{\H}<10^{-18} \Jy\sr^{-1}\cm^{2}/\H$. Yet the emission from spinning iron particles is highly polarized. The case of silicate nanoparticles looks like a possibility. But our initial estimates show that spinning emissivity from tiny silicate grains of SiO structure having large dipole moment ($\beta\approx 3.1$D per structure ) is also small if they follow the size distribution of \cite{2007ApJ...657..810D}. Thus, it requires a substantial population of silicate nanoparticles to produce higher spinning emissivity, and the processes of their formation stay unclear. However, interstellar dust has always presented us with surprises. It is clear that more work in formulating spinning dust theory of silicate nanoparticles in analogy with the existing spinning PAH theory is necessary as well as the search for the observational features that can test this hypothesis. Unlike the population of PAH, which has been proven to exist, the population of silicate nanoparticles is a hypothesis, which requires testing, e.g., through identifying particular spectral features related to the emission of these particles. This makes case (b) an interesting hypothesis that requires intensive testing.

Our present work is intended to address the issue of the polarization of MDE arising from strongly magnetic particles, in particular, free-flying nanoparticles. The discussions of MDE polarization were given for two different forms of magnetic susceptibilities given in DL99 and DH13, respectively. Although DH13 provide arguments in favor of the susceptibility that it accepted, the fit of the magnetic response to the high frequency experimental data that DH13 present is poor. It is therefore necessary to provide laboratory studies of candidate materials for the required range of frequencies, which includes the AME frequency range. We believe that this presents an important avenue for future studies related to the case (c).
Moreover, the MDE by free-flying magnetic particles was frequently ruled out as a source of AME appealing to its high expected polarization level (\citealt{LopezCaraballo:2011p508}; \citealt{2015arXiv150606660P}; \citealt{GenovaSantos:2015hc}). Our obtained results however indicate that the polarization of MDE from free-fliers is rather low, and its maximum level depends weakly on the specific form of magnetic susceptibility. Therefore, MDE from free-flying nanoparticles cannot be ruled out as a source of AME by means of polarization constraints and may be more important for AME than usually thought.

Finally, the existing problems call for revisiting ideas of non-dust sources of AME. So far, such attempts have been futile, but they were based on the incomplete knowledge of the foreground properties. For instance, the synchrotron properties may be changed by the peculiar structure of magnetic field and so the synchrotron emission spectrum. Studies of propagation as well as observational studies of the magnetic field, e.g. with new techniques suggested in \cite{2015arXiv151101537L}, may clarify the situation. We feel, however, that the case (d) is most speculative at the moment. }

\section{Summary}\label{sec:sum}
In this paper, our principal results are summarized as follows:
\begin{itemize}
\item[1] The degrees of alignment of thermally rotating silicate grains incorporated with iron nanoparticles are computed for a wide range of iron atoms per cluster $N_{\cl}$ and the fraction of atoms in clusters $f_{F}$ (characterized by $\delta_{m})$. We find that the degrees of alignment first increase with $\delta_{m}$ and reach {\it alignment saturation} at $Q_{J}\le 0.4$ and $R\le 0.07$ for $\delta_{m}\gg 1$ for the typical conditions of the ISM. The alignment saturation indicates that, without suprathermal rotation, silicate grains with magnetic inclusions are only weakly aligned.

\item[2] We have computed the degree of alignment of ferromagnetic single-domain particles, accounting for a variety of rotational damping and excitation processes. We find that the rotational damping by IR emission significantly decreases the alignment of small particles, whereas the excitation by ion collisions enhances the alignment of ultrasmall grains from the saturation level. We have also computed the alignment of Fe nanoparticles using the new susceptibility derived by DH13 and found that the degrees of alignment become independent of the specific susceptibility form when the magnetic damping is much faster than the rotational damping (i.e., $\delta_{m}\gg 1$).

\item[3] Using the computed degrees of alignment, we have predicted the polarization of MDE by free-flying iron nanoparticles. In contrast to common belief of its high polarization, we found that the polarization of MDE from free-fliers with radii of several nanometers is below $5\%$ for the typical $T_{d}=40$K. Thus, the MDE by free-flying iron nanoparticles may not be ruled out as a source of AME by appealing to the upper limits of the AME polarization.

\item[4] Spinning dust emission from iron nanoparticles with permanent magnetic moments is found to be inefficient, but its polarization level is rather high, up to $40-50\%$. The largest emissivity by iron particles is one order of magnitude lower than the emission by spinning PAHs, assuming that all Fe abundance is present in single-size nanoparticles.

\item[5] {We have calculated the polarization of spinning dust emission from PAHs for the different magnetic field strengths. The polarization is below $5\%$ for the typical physical parameters (e.g., $B\le 10\mu$G and $T_{d}\sim 60$K), which suggests that spinning dust emission by PAHs can be distinguished from spinning magnetic nanoparticles by means of the polarization signature.}

\end{itemize}

\acknowledgments
{TH thanks Huirong Yan for warm hospitality at Potsdam University, Germany. AL acknowledges NSF grant AST 1109295, support of the NSF Center for Magnetic Self-Organization, a distinguished visitor PVE/CAPES appointment at the Physics Graduate Program of the Federal University of Rio Grande do Norte and thanks the INCT INEspaço and Physics Graduate Program/UFRN for hospitality.}

\appendix
\section{A. Polarization of magnetic dust emission by free-fliers in the presence of imperfect alignment}
\subsection{A1. Coordinate system}
Let $\ahat_{1},\ahat_{2},\ahat_{3}$ be the grain principal axes where $\ahat_{1}$ is the axis of maximum moment of inertia. Let $\xhat_{J}\yhat_{J}\zhat_{J}$ are unit vectors in which $\zhat_{J}$ is parallel to $\bJ$. The observer coordinate system is defined by $\xhat\yhat\zhat$ where $\zhat$ is pointed toward the observer. Let $\xhat_{B}\yhat_{B}\zhat_{B}$ be unit vectors defined by the magnetic field in which $\zhat_{B}\| \Bv$, and the magnetic field is assumed to be in the $\yhat\zhat$ plane and makes an angle $\gamma_{B}$ with $\yhat$.  With the angles defined in the figures, we have the following:
\bea
\ahat_{1}=\cos\theta \zhat_{J} + \sin\theta(\cos\alpha \xhat_{J}+\sin\alpha \yhat_{J}),\\
\ahat_{2}=-\sin\theta \zhat_{J} + \cos\theta(\cos\alpha \xhat_{J}+\sin\alpha \yhat_{J}),\\
\ahat_{3}=[\ahat_{1}\times \ahat_{2}]=\cos\alpha \yhat_{J}-\sin\alpha \xhat_{J}.
\ena

From $J$-frame to $B$-frame, we get
\bea
\zhat_{J}=\cos\beta \zhat_{B} + \sin\beta(\cos\phi \xhat_{B}+\sin\phi \yhat_{B}),\\
\xhat_{J}=-\sin\beta \zhat_{B} + \cos\beta(\cos\phi \xhat_{B}+\sin\phi \yhat_{B}),\\
\yhat_{J}=[\ahat_{1}\times \ahat_{2}]=\cos\phi \yhat_{J}-\sin\phi \xhat_{B}.
\ena

Finally,
\bea
\zhat_{B}=\cos\gamma_{B}\xhat + \sin\gamma_{B}\zhat,~\xhat_{B}=\sin\gamma_{B}\xhat - \cos\gamma_{B}\zhat,~
\yhat_{B}=\yhat
\ena

\subsection{A2. Polarization of Magnetic Dipole Emission}\label{apdx:pol}
The polarization of dipole emission from an ensemble of grains is defined as
\bea
p=\frac{I_{x}-I_{y}}{I_{x}+I_{y}},\label{eq:p_IxIy}
\ena
where $I_{x}$ and $I_{y}$ are the radiation intensity with the electric field $\bE$ parallel to $\xhat$ and $\yhat$ in the plane of the sky. 

For a single grain, Equation (\ref{eq:p_IxIy}) can be rewritten as
\bea
p=\frac{C_{x}-C_{y}}{C_{x}+C_{y}},~~~C_{x,y}=C_{x,y}^{\ed}+C_{x,y}^{\md},\label{eq:pol}
\ena
where $C_{x}=C_{\abs}(\bE\| \xhat)$, and $C_{y}=C_{\abs}(\bE\| \yhat)$ are total absorbtion cross-section due to dielectric susceptibility and magnetic permeability, which arises from the interaction of the component $\bE$ with electric dipoles and $\bB$ with magnetic moments. The former is dominant for high frequencies (UV, optical and IR wavelengths) while the latter is important for low frequencies (cm and submm).

%Using the definition of cross-section, we can write
%\bea
%C_{x} = (\xhat.\ahat_{1})^{2}C_{\abs}(E\| \ahat_{1}) + (\xhat.\ahat_{2})^{2}C_{\abs}(E\| \ahat_{2}) + (\xhat.\ahat_{3})^{2}C_{\abs}(E\| \ahat_{3}),\label{eq:Cx}\\
%C_{y} = (\yhat.\ahat_{1})^{2}C_{\abs}(E\| \ahat_{1}) + (\yhat.\ba_{2})^{2}C_{\abs}(E\| \ahat_{2}) + (\yhat.\ahat_{3})^{2}C_{\abs}(E\| \ahat_{3}),\label{eq:Cy}
%\ena

% Why a1<a2<a3 then L1>L2>L3. What are physical justification? Let consider a sphere where L1=L2=L3. Since we approximate an ellipsoid to an equivalent sphere of volume V, we need to introduce some correction geometrical factor. The long axis has larger length than the equivalent radius $a_{eff}$. So, the geometrical factor must be smaller than 1. The short axis has smaller length than the equivalent radius, so, the geometric factor must be larger to bring the ellipsoid to the sphere. Thus, the geometrical factors act to transform an ellipsoid to an equivalent sphere, where the key is we approximate an irregular grain by a sphere of equivalent volume adopted for easy calculations.

% Magnetic emission at high frequency nu>1GHz cannot be calculated using DL99 results. I must use DH13 results.

For an ellipsoidal grain of semiaxes $a_{1}\le a_{2}\le a_{3}$ with geometrical factors $L_{1}\ge L_{2}\ge L_{3}$, the electric cross section for the incident electric field parallel to the semi-axis $\be \equiv \ahat_{i}$, and magnetic absorption cross-section for the magnetic field parallel to $\hhat$ axis perpendicular to $\be$ are
\bea
C_{\abs}^{\ed}(\bE\| \ehat) \approx V\frac{\omega}{c}\left[\frac{\epsilon_{2}}{|1+L_{e}(\epsilon-1)|^{2}} \right],~~
C_{\abs}^{\md}(\bE\| \ehat;\bH\| \hhat)\approx V\frac{\omega}{c}\left[\frac{\mu_{2}}{|1+L_{h}(\mu-1)|^{2}} \right],\label{eq:Cabs}
\ena
where $\epsilon$ and $\mu$ are dielectric polarizability and isotropic magnetic permeability (see \citealt{1983asls.book.....B}). So, the cross-section for $\bE$ (or $\bH$) along the short axis (larger $L_{j}$) is smaller than the cross-section for $\bE$ along the long axis. 
%It means if E is parallel to e = a1 axis, then H is parallel to the plane a2a3, so h-axis can be a2 and a3. This notation is just to indicate the difference between E and H components.

%This relation is equivalent to the choice of $\xhat\yhat\zhat$ such that the magnetic field $\Bv$ lies in the plane $\xhat\zhat$ for which the projection of $\Bv$ in the sky plane is directed along $\xhat$.

For a simple situation in which the rotation is along the short axis $\ahat_{1}\|\bJ\| \xhat$, and both axes are directed along the magnetic field in the sky plane $\xhat\yhat$, DL99 got the polarization $p=(C_{x}-C_{y})/(C_{x}+C_{y})$ with
\bea
C_{x} =V\frac{\omega}{c}\left[\frac{\epsilon_{2}}{|1+L_{1}(\epsilon-1)|^{2}}+\frac{1}{2}\left(\frac{\mu_{\perp,2}}{|1+L_{2}(\mu_{\perp}-1)|^{2}}\right)\right],\label{eq:Ch}\\
C_{y} = V\frac{\omega}{c}\left[\frac{1}{2}\left(\frac{\epsilon_{2}}{|1+L_{2}(\epsilon-1)|^{2}}+ \frac{\epsilon_{2}}{|1+L_{3}(\epsilon-1)|^{2}} \right)+\frac{\mu_{\perp,2}}{|1+L_{1}(\mu_{\perp}-1)|^{2}}\right],\label{eq:Ce}
\ena
where $\yhat$ is the same as $e-$axis, and $\xhat$ is the same as ${\bf h}-$axis in DL99.  Note that $C_{x}$ is defined such that incident electric field $\bE$ is along $\xhat$, which directly means $\bH$ is directed along $\yhat$ since the wave plane is $\xhat\yhat$ with propagation direction $\zhat$. Similarly is $C_{y}$ with $\bE\| \yhat$ and ${\bf H}\| \xhat\equiv \ahat_{1}$. 

%That is the reason why there is the term $C_{\abs}^{\md}(H\| \ahat_{1})$ in $C_{e}\equiv C_{y}(E\| \yhat)$. This relation is valid for perfect alignment. Here the factor $1/2$ accounts for the average of $\cos^{2}$ and $\sin^{2}$ over fast rotation around the short axis.

When the imperfect alignment of grains with the magnetic field is considered, the polarization (Eq. \ref{eq:pol}) for oblate grains becomes (see \citealt{1985ApJ...290..211L}; RL99)
\bea
C_{x}-C_{y}=C_{\pol}R\cos^{2}\gamma_{B},~~C_{x}+C_{y}={2C_{\rm avg} +(RC_{\pol}\cos^{2}\gamma/3)(2/\cos^{2}\gamma_{B}-3)},
\ena
where $C_{\rm avg}=(2C_{\perp} + C_{\|})/3$, and $C_{\pol}=C_{\perp}-C_{\|}$, including both electric dipole and magnetic dipole extinction. In this equation, 
\bea
C_{\|} =V\frac{\omega}{c}\left[ \frac{\epsilon_{2}}{|1+L_{1}(\epsilon-1)|^{2}}+\frac{1}{2}\frac{\mu_{\perp,2}}{|1+L_{2}(\mu_{\perp}-1)|^{2}}\right],\\
C_{\perp}=V\frac{\omega}{c}\left[\frac{\epsilon_{2}}{|1+L_{2}(\epsilon-1)|^{2}}+\frac{\mu_{\perp,2}}{|1+L_{1}(\mu_{\perp}-1)|^{2}}\right],
\ena
for $L_{2}=L_{3}$.

The dielectric function for ferromagnetic particles is taken to be
\bea
 \epsilon(\omega) = \frac{i(\omega_{p}\tau)^{2}}{\omega \tau-i(\omega\tau)^{2}},
 \ena
where $\tau=3.8\times 10^{-14}s$ and $\omega_{p}\tau\sim 200$.

\section{B.Rotational Damping and Excitation Coefficients for Magnetic Particles}
\subsection{B1.Grain geometry}
We consider oblate spheroidal grains with moments of inertia $I_{1}>I_{2}=I_{3}$ along the grain principal axes denoted by $\hat{\ba}_{1}$, $\hat{\ba}_{2}$ and $\hat{\ba}_{3}$. Let $I_{\|}=I_{1}$ and $I_{\perp}=I_{2}=I_{3}$. They take the following forms: 
\bea
I_{\|}=\frac{2}{5}Ma_{2}^{2}=\frac{8\pi}{15}\rho a_{1}a_{2}^{4},~I_{\perp}=\frac{4\pi}{15}\rho a_{2}^{2}a_{1}\left(a_{1}^{2}+a_{2}^{2}\right),\label{eq:Iparperp}
\ena
where $a_{1}$ and $a_{2}=a_{3}$ are the lengths of the semiminor and semimajor axes of the oblate spheroid with axial ratio $s= a_{1}/a_{2}<1$, and $\rho$ is the grain material density. 

The grain size $a$ is defined as the radius of an equivalent sphere of the same volume, which is given by
\bea
a=\left(\frac{3}{4\pi} (4\pi/3) a_{1}a_{2}^{2}\right)^{1/3}=a_{2}s^{1/3}.\label{eq:aeff}
\ena

\subsection{B2. Rotational damping and excitation coefficients}\label{apdx:FGcoeff}
We follow the definitions of rotational damping $F$ and excitation coefficients $G$ from \cite{1998ApJ...508..157D}. The dimensionless damping coefficient for the $j$ process, $F_{j}$, is defined as the ratio of the damping rate induced by that process to that induced by the collisions of gas species, $\tau_{\H}^{-1}$, assuming that the gas consists of purely atomic hydrogen:
\bea
F_j=\left(-\frac{d\omega}{\omega dt}\right)_{j}\left(\frac{1}{\tau_{\H}^{-1}}\right)\label{eq:F}
\ena
and the excitation coefficient is defined as
\bea
G_j=\left(\frac{I d\omega^{2}}{2dt}\right)_{j}\left(\frac{\tau_{\H}}{\kB T_{\rm gas}}\right),\label{eq:G}
\ena
where $j$=n, i, p and IR denote the grain collisions with neutrals and ions, plasma-grain interactions, and the IR emission, $\left(I d\omega^{2}/2dt\right)_{j}$ is the rate of increase of kinetic energy for rotation along the axis that has moment of inertia $I$ due to the excitation process $j$. For an uncharged grain in a gas of purely atomic hydrogen, $F_{n}=1$ and $G_{n}=1$.

The thermal angular velocity of a grain around its symmetry axis (of rotational energy $k_{\B}T_{\gas}$) is
\bea
\omega_{T}=\left(\frac{2k_{B}T_{\gas}}{I_{\|}} \right)^{1/2}\simeq 3.3\times 10^{5}a_{-5}^{-5/2}s^{2/3}\hat{\rho}^{-1}\hat{T}_{\gas}^{1/2} \s^{-1},\label{eq:omegaT}
\ena
where $\hat{\rho}=\rho/3\g\cm^{-3}$. 

To calculate the damping and excitation coefficients for wobbling grains, we follow the same approach as in HDL10, where the parallel components $F_{j,\|}$ and $G_{j,\|}$, and perpendicular components $F_{j,\bot}$ and $G_{j,\bot}$ with respect to $\ahat_{1}$ are computed using the general definitions (Equations \ref{eq:F} and \ref{eq:G}). The only modification is the moments of inertia $I_{\|}$ and $I_{\perp}$ and $\tau_{\H,\|}$ and $\tau_{\H,\perp}$ for an oblate spheroid instead of those for disk-like grains in HDL10.

For example, the characteristic damping times of an oblate spheroidal grain for rotation along the directions parallel and perpendicular to the grain symmetry axis $\ahat_{1}$ are respectively given by (\citealt{1997MNRAS.288..609L})
\bea
\tau_{\H,\|}=\frac{3I_{\|}}{4\sqrt{\pi}n_{\H}m_{\H}v_{\th}a_{2}^{4}\Gamma_{\|}},~~\tau_{\H,\perp}=\frac{3I_{\perp}}{4\sqrt{\pi}n_{\H}m_{\H}v_{\th}a_{2}^{4}\Gamma_{\perp}},\label{eq:tauHxy}
\ena
where $\tau_{\H,\|}\equiv\tau_{\H,z}, \tau_{\H,\perp}\equiv \tau_{\H,y}=\tau_{\H,x}$ with $z$ the grain symmetry axis, and $x$ and $y$ being the axes perpendicular to the symmetry axis. In the above equation, $n_{\H}$ is the gas density, $m_{\H}$ is the hydrogen mass, $v_{\th}=(2k_{\B}T_{\gas}/m_{\H})^{1/2}$ is the thermal speed of hydrogen, and the $\Gamma_{\|}$ and $\Gamma_{\perp}$ are geometrical factors (\citealt{1993ApJ...418..287R}).

For the typical parameters of the ISM, Equations (\ref{eq:tauHxy}) become
\bea
\tau_{\H,\|}\simeq 6.58\times 10^{4} \hat{\rho}\hat{s}^{2/3} a_{-5}
\hat{n}_{\gas}^{-1}\hat{T}_{\gas}^{-1/2} \Gamma_{\|}^{-1}\yr,~~~
\tau_{\H,\perp}\simeq 4.11\times 10^{4}\hat{\rho}\hat{s}^{2/3}\left(\frac{1+s^{2}}{1.25}\right) a_{-5}\hat{n}_{\gas}^{-1}\hat{T}_{\gas}^{-1/2} \Gamma_{\perp}^{-1}\yr,
\ena
where $\hat{s}=s/0.5$.

Likewise, the characteristic damping times due to the electric dipole emission from HDL10 can be rewritten as
\bea
\tau_{\ed,\|}=\frac{3I_{\|}c^{3}}{6k_{\B}T_{\gas}\mu_{\perp}^{2}},~~
\tau_{\ed,\perp}=\frac{3I_{\perp}c^{3}}{6k_{\B}T_{\gas}\left(\mu_{\perp}^{2}/2+\mu_{\|}^{2}\right)},\label{eq:tauedxy}
\ena
where $\mu_{\|}$ and $\mu_{\perp}$ are the components of the electric dipole moment $\bmu$ parallel and perpendicular to the grain symmetry axis. Here we assume an isotropic distribution of $\bmu$, which corresponds to  $\mu_{\|}^{2}=\mu_{\perp}^{2}/2=\mu^{2}/3$ where $\mu^{2}$ is given by Equation (11) in DL98.

The geometrical factors in Equations (\ref{eq:tauHxy}) are given by
\bea
\Gamma_{\|}=\frac{3}{16}\left[3+4(1-e^2)g(e)-e^{-2}(1-(1-e^2)^2)g(e)\right],\label{eq:Gam_par}\\
\Gamma_{\perp}=\frac{3}{32}\left[7-e^2+(1-e^2)^{2}g(e)+(1-2e^2)(1+e^{-2}
[1-(1-e^2)^2)g(e)])\right],\label{eq:Gam_per}
\ena
where $e=\sqrt{1-s^{2}}$ and 
\bea
g_{e}=\frac{1}{2e}\ln\left(\frac{1+e}{1-e}\right).\label{eq:ge}
\ena

\subsection{B3. Passing ion-permanent magnetic dipole interactions}\label{apd:plasma}

Let consider an ion of charge $q$ passing a magnetic grain in $\xhat$ direction with velocity $v$. The magnetic field produced by the moving charge at the grain position is
\bea
{\Bv} = \frac{q\bv/c\times {\bf r}}{r^{3}},
\ena
where $\br =x\xhat + y\yhat$ is the distance from the moving charge to the grain position.

It is easy to see that the magnetic torque is zero when the magnetic moment of the magnetic particle ${\bf m}$ is directed along the $\zhat$ axis, and is nonzero when ${\bf m}$ lies in the plane $\xhat\yhat$. Let's assume that ${\bf m}$ is directed along $\yhat$ for simplicity. The torque acting on the dipole is then equal to
\bea
\Gamma = {\bf m}\times {\bf B} = \frac{qmvy}{cr^{3}}\xhat.
\ena

For the impact approximation, the ion is assumed to be moving along a straight trajectory (e.g., $\xhat$ axis) at constant $y$. The angular momentum impulse induced by the entire ion trajectory is given by
\bea
\delta J = \int_{-\infty}^{\infty}\Gamma dt=\int_{-\infty}^{\infty} \Gamma \frac{dx}{v} = \int_{-\infty}^{\infty}\frac{qm ydx}{c(x^{2}+y^{2})^{3/2}}=\frac{2qm}{cy}.
\ena

Therefore, the increase of rotational energy per second is 
\bea
\frac{(\Delta J)^{2}}{\Delta t} = \dot{N}_{ion} (\delta J)^{2},
\ena
where $\dot{N}_{ion}=n_{i}v 4\pi v^{2}Ze^{-\alpha v^{2}}dv  2\pi y dy$ is the rate of incident ions.

Adopting a Maxwellian velocity distribution for ions, we obtain
\bea
\frac{(\Delta J)^{2}}{\Delta t} &=& \int_{0}^{\infty}\int_{0}^{b_{\max}} n_{i}v4\pi v^{2}Ze^{-\alpha v^{2}} dv \frac{4q^{2}m^{2}}{c^{2}}\frac{dy}{y}=8\pi n_{i}\langle v\rangle \left(\frac{qm}{c}\right)^{2}\ln\left(\frac{b_{\max}}{a} \right)
\ena
where $\langle v\rangle = \left(8k_{B}T_{gas}/\pi m_{i} \right)^{1/2}$ is the mean ion velocity.

As usual, the excitation coefficient is defined as
\bea
G_{p}= \frac{(\Delta J)^{2}}{\Delta t}: \frac{\left(I_{\|}\omega_{T}\right)^{2}}{\tau_{H,\|}}=8\pi n_{i}\langle v\rangle \left(\frac{qm}{c}\right)^{2}\ln\left(\frac{b_{\max}}{a} \right)\times \frac{\tau_{H,\|}}{I_{\|}k_{\B}T_{\gas}}.
\ena

Using $m= (4\pi M_{s}/10^{3}{\rm G}) a_{-7} 10^{-18}$ c.g.s, and plugging numerical parameters into the above equation, it yields $G_{p}\sim 10^{-10}$. This indicates that the excitation by passing ions is negligible.

\section{C. Second-order integrator for Langevin Equation}\label{apdx:LEsolver2}
\subsection{C1. Integrator algorithms}

Earlier works on numerical calculations of magnetic alignment used the first-order integrator (Euler-Maruyama algorithm) to solve the Langevin equation. In order to account for high accuracy of alignment degree relevant for polarization, we implement second-order integrator. As shown in \cite{VandenEijnden:2006gp}, the accuracy of the second-order integrator is more than one order of magnitude of the first-order.

By setting $j=J'$ and omitting the primes in $dt'$ and $dw'$, Equation (\ref{eq:dJp_dt}) can be rewritten as
\bea
dj_{i} =-\left(\gamma_{i} j_{i} + \gamma_{\ed}{j_{i}^{3}}\right) dt+  \sigma_{ii} dw(t),\label{eq:dJdt_W}
\ena
where $\gamma_{i}=1/\tau'_{\gas,{\eff}} +\delta_{m}(1-\delta_{zi})$, $\sigma_{ii} = \sqrt{B'_{ii}}$ in our case. The momentum component $i$ at $t+h$ is obtained by integrating the above equation from $t$ to $t+h$:
\bea
j(t+h) = j(t)  -\gamma \int_{t}^{t+h}j(s)ds- \gamma_{\ed}\int_{t}^{t+h} j(s)^{3}ds + \sigma [w(t+h)-w(t)],\label{eq:jnp1}
\ena
where the index $i$ has been omitted.

Note that $w(t+h)-w(t) \,{\buildrel d \over =}\, \sqrt{h}\zeta$, where $ \,{\buildrel d \over =}\,$ denotes the equality in distribution, and $\zeta$ is a Gaussian random variable with mean zero and unit variance. In the first-order approximation, Equation (\ref{eq:dJdt_W}) yields the solution at iterative step $n+1$ as follows:
\bea
j(t+h) &=&j(t) - \left( \gamma j(t) +\gamma_{\ed}j(t)^{3}\right)h+\sqrt{h}\sigma \zeta.
\ena
which is well-known Euler-Maruyama algorithm.

To achieve higher accuracy, following \cite{VandenEijnden:2006gp}, we write $j(s)$ in Equation (\ref{eq:jnp1}) as the follow: 
\bea
j(s)&=&j(t)-\int_{t}^{s} \gamma j(u)du+\sigma[w(s)-w(t)]- \int_{t}^{s}\gamma_{\ed}j(u)^{3}du,\\
\ena
where $t<s<t+h$.

Using $\langle w^{2}\rangle = h$, we get $j(u) = j(t) + O(h^{1/2})$ and $
\int_{t}^{s}j(u)^{3}du = j(t)^{3}(s-t) + O(h^{3}/2)$. Thus, the above equation becomes
\bea
j(s)&=&j(t) -\gamma j(t) (s-t) + \sigma[w(s) − w(t)] -\gamma_{\ed}j(t)^{3}(s-t) + O(h^{3}/2),
\ena

Using the fact that $w(s)-w(t) \,{\buildrel d \over =}\,(s-t)^{1/2}\xi$ , we have $
\int_{t}^{t+h}[w(s)-w(t)]ds \,{\buildrel d \over =}\,h^{3/2}\left(\xi^{2}/2+\eta^{2}/2\sqrt{3}\right)$, and
 \bea
\int_{t}^{t+h}j(s)ds=j(t)h-\gamma j(t) h^{2}/2 +\sigma h^{3/2} g(\xi_{n},\eta_{n})−\gamma_{\ed}j (t)^{3} h^{2}/2 +O(h^{5/2} ),\label{eq:jsp1}\\
\int_{t}^{t+h}j(s)^{3} ds=j (t)^{3}h-3\gamma_{\ed}j(t)^{3}\left(\gamma +\gamma_{\ed}j(t)^{2}\right) h^{2}/2 + 3j (t)^{2}\sigma h^{3/2} g(\xi_{n},\eta_{n}) + O(h^{5/2} ), \label{eq:j3sp1}
\ena
where $\eta_{n}$ and $\xi_{n}$ are independent Gaussian variables with zero mean and unit variance, and $g(\xi_{n} , \eta_{n} ) = \xi_{n}^{2} /2 + \eta_{n}^{2} /2\sqrt{3}$ 

Let $j_{n}\equiv j(t)$ and $j_{n+1}\equiv j(t + h)$. Plugging Equation (\ref{eq:jsp1}) and (\ref{eq:j3sp1}) into (\ref{eq:jnp1}) we obtain the following to the second order:
\bea
{j}_{i;n+1} &=& j_{i;n}  -  \gamma_{i}{j}_{i;n}h+\sigma_{ii} \sqrt{h}{\zeta}_{n}- \gamma_{i}\mathcal{A}_{i;n}-\gamma_{\ed}\mathcal{B}_{i;n},
\ena
where
\bea
\mathcal{A}_{i;n}&=& -\frac{h^{2}}{2} \gamma_{i} j_{i;n}+\sigma_{ii} h^{3/2}g(\xi_{n},\eta_{n})-\gamma_{\ed}j_{i;n}^{3}\frac{h^{2}}{2},\\
\mathcal{B}_{i;n}&=&j_{i;n}^{3}h - 3j_{i;n}^{3}\left(\gamma_{i}+\gamma_{\ed}j_{i;n}^{2}\right)\frac{h^{2}}{2}+3j_{i;n}^{2}\sigma_{ii} h^{3/2}g(\xi_{n},\eta_{n})
\ena

The advantage of second-order integrator is that it allows us to achieve the comparable accuracy as the first-order, but with larger timestep $h$, which can save significant computing time.

%\subsection{C2. Numerical tests of second-order vs. first-order integrator}
%Here, we compare numerical errors for our problems using first- and second-order integrator. We perform $N$ simulations in which each simulations corresponds to an integration of LE for $T=100\tau_{gas}$. The mean value over the simulations is denoted by $\overline{Q_{J}}$, mean error is $\delta Q=Q-Q_{0}$, and rms value $\sigma_{Q}=\overline{(\delta Q)^{2}}^{1/2}$, where $Q_{0}$ is the exact value for sphere. 

%--------------adding references-----------------------------------
%\bibliographystyle{/Users/thiemhoang/Dropbox/Papers2/apj}
% or other styles: mcbride,plain, abbrv, acm, alpha, apalike, apj
%\bibliography{/Users/thiemhoang/Dropbox/Papers2/cites_paperApJ,/Users/thiemhoang/Dropbox/Papers2/cites_Books}

\bibliography{ms.bbl}

%\bibitem[\protect\citeauthoryear{Lazarian \& Hoang}{Lazarian \&
%  Hoang}{2008}]{Lazarian:2008fw}
%Lazarian, A., Andersson, B-G, \& Hoang, T. 2014, in Polarization of stars and planetary systems, eds. L. Kolokolova, J. Hough, A.-Ch. Levasseur-Regourd (New York:
%Cambridge University Press)

\end{document}